\RequirePackage{lineno}
\pdfoutput=1 
\documentclass[aps,prd,twocolumn,showpacs,superscriptaddress,groupedaddress,floatfix]{revtex4-2}  
\usepackage{graphicx}  
\usepackage{bm}        
\usepackage{amssymb}   
\usepackage{amsmath}   
\usepackage{amsfonts}
\usepackage{color}

\begin{document}

\title{Comparison of transverse single-spin asymmetries for forward $\pi^{0}$ production in polarized $pp$, $p\rm{Al}$ and $p\rm{Au}$ collisions at nucleon pair c.m. energy $\sqrt{s_{\mathrm{NN}}}= 200$ GeV}

\author{
J.~Adam$^{6}$,
L.~Adamczyk$^{2}$,
J.~R.~Adams$^{39}$,
J.~K.~Adkins$^{30}$,
G.~Agakishiev$^{28}$,
M.~M.~Aggarwal$^{41}$,
Z.~Ahammed$^{61}$,
I.~Alekseev$^{3,35}$,
D.~M.~Anderson$^{55}$,
A.~Aparin$^{28}$
% ...
}

% ...
\address{$^{2}$AGH University of Science and Technology, FPACS, Cracow 30-059, Poland}
\address{$^{3}$Alikhanov Institute for Theoretical and Experimental Physics NRC "Kurchatov Institute", Moscow 117218, Russia}
% ...
\address{$^{6}$Brookhaven National Laboratory, Upton, New York 11973}
% ...
\address{$^{28}$Joint Institute for Nuclear Research, Dubna 141 980, Russia}
% ...
\address{$^{30}$University of Kentucky, Lexington, Kentucky 40506-0055}
% ...
\address{$^{35}$National Research Nuclear University MEPhI, Moscow 115409, Russia}
% ...
\address{$^{39}$Ohio State University, Columbus, Ohio 43210}
% ...
\address{$^{41}$Panjab University, Chandigarh 160014, India}
% ...
\address{$^{55}$Texas A\&M University, College Station, Texas 77843}
% ...
\address{$^{61}$Variable Energy Cyclotron Centre, Kolkata 700064, India}
% ...

%

\collaboration{STAR Collaboration}\noaffiliation


\begin{abstract}
The STAR Collaboration reports a measurement of the transverse single-spin asymmetries, $A_{N}$, for neutral pions produced in polarized proton collisions with protons ($pp$), with aluminum nuclei ($p\rm{Al}$) and with gold nuclei ($p\rm{Au}$) at a nucleon-nucleon center-of-mass energy of 200 GeV. Neutral pions are observed in the forward direction relative to the transversely polarized proton beam, in the pseudo-rapidity region $2.7<\eta<3.8$. Results are presented for $\pi^0$s observed in the STAR FMS electromagnetic calorimeter in narrow Feynman x ($x_F$) and transverse momentum ($p_T$) bins, spanning the range $0.17<x_F<0.81$ and $1.7<p_{T}<6.0$ GeV/$c$. For fixed $x_F<0.47$, the asymmetries are found to rise with increasing transverse momentum. For larger $x_F$, the asymmetry flattens or falls as ${p_T}$ increases. Parametrizing the ratio $r(A) \equiv A_N(pA)/A_N(pp)=A^P$ over the kinematic range, the ratio $r(A)$ is found to depend only weakly on $A$, with ${\langle}P{\rangle} = -0.027 \pm 0.005$. No significant difference in $P$ is observed between the low-$p_T$ region, $p_T<2.5$ GeV/$c$, where gluon saturation effects may play a role, and the high-$p_T$ region, $p_T>2.5$ GeV/$c$. It is further observed that the value of $A_N$ is significantly larger for events with a large-$p_T$ isolated $\pi^0$ than for events with a non-isolated $\pi^0$ accompanied by additional jet-like fragments. The nuclear dependence $r(A)$ is similar for isolated and non-isolated $\pi^0$ events.
\end{abstract}

\maketitle

\section{Introduction}

The measurements and the evolving interpretations of transverse single-spin asymmetries for forward pion production in high energy $pp$ collisions have a rich 
history ~\cite{Klem:1976ui,Saroff:1989gn,Adams:1991rw,Adams:1991cs,Arsene:2008aa,Adare:2013ekj,Abelev:2008af}. 
These measurements guided the development of Quantum Chromo-Dynamics (QCD) based models that incorporated quark helicity conservation, QCD factorization, the nature of initial state parton motion or angular momentum, and the dynamics of fragmentation within the scattering processes for polarized protons.
The new transverse asymmetry measurements,
presented here, again challenge aspects of current models for the application of  QCD to the spin dependence of cross sections.
The $\pi^0$ single-spin asymmetry, $A_N$, is measured as a function of pion kinematics for collisions between polarized protons and protons ($pp$), aluminum nuclei ($p\rm{Al}$) and gold nuclei ($p\rm{Au}$). Because $A_N$ for this process is expected to be very sensitive to the QCD fields in the vicinity of a struck quark, the nuclear dependence of $A_N$ should be sensitive to phenomena that modify the local fields, for example, gluon saturation effects.

This analysis presents the dependence of $A_N$ in the forward $\pi^0$ production process,
$p^\uparrow + p( \rm{or}\: A) \rightarrow \pi^0 + X$.
It is useful to first define a simple azimuthal angle-dependent asymmetry, $a_N(x_F,p_T,\phi)$, as the ratio of the difference in cross section for the two proton transverse spin states, $\sigma^\uparrow$ and $\sigma^\downarrow$, to the sum of those cross sections for a pion produced at $x_F$ (Feynman X) and $p_T$ (transverse momentum),
\begin{linenomath}
\begin{align}
\label{eq:andef}
a_N(x_F,p_T,\phi)=\frac{\sigma^\uparrow(x_F,p_T,\phi) -\sigma^\downarrow(x_F,p_T,\phi)}
{\sigma^\uparrow(x_F,p_T,\phi) +\sigma^\downarrow(x_F,p_T,\phi)}\\
\label{eq:Andef}
=A_N(x_F,p_T)\cos\phi.
\end{align}
\end{linenomath}
The three components of pion momentum are specified with coordinates $x_F$, $p_T$ and $\phi$. 
The dependence of the pion differential cross section on transverse spin, expressed as the pion momentum dependent asymmetry  $a_N(x_F,p_T,\phi)$ and
the transverse single-spin asymmetry, $A_N(x_F,p_T)$, are defined in terms of the simple asymmetry accordingly, (Eq.~\ref{eq:Andef}).
Referring to a right-handed coordinate system,  an initial state polarized proton is referred to as spin ``up" if it has a positive spin projection along the $y$ axis while proton momentum is along the $z$ axis. This polarized proton collides with an unpolarized proton or nucleus traveling along the $-z$ axis.  A forward pion has a positive longitudinal component of momentum $p^\pi_L$, given by a positive fraction 
$x_F=2\frac{p^\pi_L}{\sqrt{s}}$ of the polarized proton momentum.  The angle $\phi$ is the pion azimuthal angle about the $z$ axis measured from the $x$ axis positive direction. 
Equation \ref{eq:Andef} defines $A_N(x_F,p_T)$ in terms of cross sections, which are differential in $x_F$, $p_T$ and $\phi$, with superscript arrows indicating the spin directions up or down, respectively. Symmetry requires that the $\phi$ dependence be proportional to $\cos\phi$.

\section{The Relation between Scattering with Longitudinal and Transverse Polarization}
The unique features of the spin dependence of scattering from quarks or gluons in transversely polarized protons are best understood when contrasted with scattering of partons in longitudinally (helicity) polarized protons. 
For a longitudinally polarized Dirac fermion, the dependence of cross section on the initial state spin is connected to helicity
conservation.
For a relativistic electron or quark, the absorption or emission of a virtual photon (or similarly a gluon) cannot flip the helicity of a relativistic fermion. 
However, in a one dimensional scattering example, where a virtual photon is in a particular helicity state, and is absorbed by a free quark at rest, the longitudinal spin component of the quark must flip as one unit of photon spin is absorbed by the quark, changing the 
struck quark spin by one unit. Such a photon can be absorbed by only one of the two possible initial quark spin states, so cross sections thus can depend on initial state quark spin component along the final state direction or on the final state helicity.
But with absorption from a transversely polarized quark, where transverse spin states are composed of equal magnitude combinations of the two helicity states, the cross section is the same for either transverse spin state.
For scattering between small mass electrons and quarks, 
this generalizes to a cross section, 
that depends on Dirac fermion helicities but not on their transverse spins. 
Any cross section dependence on transverse spin is associated with the negligibly small helicity flip amplitudes.

In the original quark model, where the spin of a polarized proton was attributed to the polarized quarks, it was clear that the longitudinal polarization of these quarks could be observed by the double helicity measurements in scattering between protons and electrons. 
Because deep inelastic scattering cross sections were most sensitive to the {\it{up}} quarks, due to their larger electric charge, it was a very early  prediction of the quark model that the 
longitudinal polarization of {\it{up}} quarks within the polarized proton could be observed by measuring the dependence of the lepton-proton cross section on the proton and lepton longitudinal spins~\cite{Bjorken:1969mm}.

The longitudinal double spin lepton-proton scattering measurements provided the mechanism for the first measurements of quark momentum dependent longitudinally polarized quark distributions in a longitudinally polarized proton~\cite{Alguard:1976bm, Ashman:1987hv}.  Similar longitudinal double spin proton-proton cross sections depended upon the longitudinal polarization of partons, including gluons. Measurements and analysis of longitudinally polarized protons remain an important topic for the STAR experiment, to constrain longitudinal polarization densities of partons in the proton. Global analyses of many experiments~\cite{deFlorian:2009vb, Nocera:2014gqa,deFlorian:2014yva} have integrated the experimental results.

In a frame where the proton was highly relativistic, where each quark momentum was nearly parallel to the proton momentum, the cross section did depend directly on the helicity of the struck quark. The cross section associated with a longitudinally polarized, nearly free, quark was calculable from hard scattering in helicity conserving perturbative processes.  
The longitudinal double spin asymmetry was then sensitive to the longitudinally polarized struck  quark in the longitudinally polarized proton.
However, the scattering cross section for such a quark did not depend on the components of its spin measured along a transverse axis. 
Such a dependence would have been associated with
the parton flipping helicity as it interacted, by absorbing or emitting a photon or gluon. 
Because the quark helicity-flip amplitude was vanishingly small at high energies, early predictions, that the transverse spin dependence of the quark scattering process should vanish at high energy, implied that $A_N$ should be small for high energy collisions~\cite{Kane:1978nd}. 
Transverse spin dependence of cross sections are known to be further
suppressed because such dependencies required an interference between helicity amplitudes with  different phases.
Such a phase-shifted amplitude is not present in the hard scattering part of leading twist perturbative QCD (pQCD) processes.

From the above discussion, it is clear that the helicity conserving hard parton amplitudes, which are apparently dominant in the unpolarized cross sections, imply calculable sensitivity of the parton cross sections to parton helicity. This leads to longitudinal asymmetries, reflecting the polarization of partons in the proton. In contrast, the corresponding hard isolated amplitudes are insensitive to  the transverse spin of the partons. The large transverse spin asymmetries, $A_N$, in $pp$ collisions reveal physics beyond that of hard isolated parton scattering. 

\section{Mechanisms for non-zero Transverse Asymmetry}
The measurement of transverse spin asymmetries is sensitive 
to effects that are very different than the
physics responsible for longitudinal asymmetries.
The traditional pQCD calculations for hard scattering from protons relied on collinear factorization~\cite{Collins:1989gx}, where all parton momenta were characterized as propagating parallel to the parent proton momentum. 
Within this framework, the transverse spin dependence was limited by the suppression of hard scattering helicity-flip amplitudes.   
But more nuanced pictures of scattering of quarks in a transversely polarized proton have emerged, utilizing parton density distributions that characterize both transverse and longitudinal components of parton momentum. 
With such a parton density distribution, the initial state parton motion need not be parallel to the proton momentum, meaning that a helicity frame for the proton may not completely align with the helicity frame of the quark.

A transverse momentum offset of $\vec{k}_T$, representing the average transverse momentum of the initial/final state quark relative to the initial/final state parent hadron, respectively, is added to the transverse momentum $\vec{P}_T$ from the hard scattering process to form the observed pion transverse momentum,
$p^\pi_T = |\vec{P}_T+\vec{k}_T|$. 
So while the quark scattering cross section has little direct dependence on the transverse spin of the quark, the pion production cross section can depend on the transverse spin of the proton through initial and final state interactions leading to non-zero $\vec{k}_T$.
If this bias of $\vec{k}_T$ is correlated with the transverse spin of the proton, then non-zero $A_N$ will result. 
This kind of proton spin dependence of the observed pion cross section is amplified by the extreme ${p}_T$ dependence of the hard pion cross section. 

The general expectation that the pion $A_N$ should fall with increasing $p_T$ for $p_T$ above a nominal QCD momentum scale can be demonstrated in a simple model. 
If one assumes that  the forward hard scattering cross section of a quark, with momentum fraction $x$, falls with increasing transverse momentum, $p_T$, by a power law form with power $N$, then
\begin{linenomath}
\begin{equation} 
\frac{d\sigma}{dp_T}\propto p_T^{-N},
\label{Eq:powN}
\end{equation}
\end{linenomath}
where $p_T=|\vec{P}_T|$. 
If the scattered quark acquires transverse 
momentum $\vec{k}_T=\pm{k}_T \hat{x}$ 
from initial or final state interactions that is correlated with the polarized proton spin in the $\pm\hat{y}$ directions, then we see that $A_N$ will also fall with increasing $p_T$. Assuming the hard scattering transverse momentum is much greater than the initial state or final state transverse momentum ($p_T \gg k_T$), then the difference in cross section when $p^\pi_T$ is measured along the $\pm\hat{x}$ direction leads to $A_N$ as in Eq.~\ref{Eq:leftright}. If we assume a 
cross section form for $p_T$, as in Eq.~\ref{Eq:powN}, expressing $A_N$ as a left-right asymmetry, we have
\begin{linenomath}
\begin{equation}
\begin{aligned}
A_N(x_F,p_T)&=\frac{\sigma^\uparrow(x_F,p_T,0) -\sigma^\uparrow(x_F,p_T,\pi)}{\sigma^\uparrow(x_F,p_T,0) +\sigma^\uparrow(x_F,p_T,\pi)}\\
&\simeq \frac{(p^\pi_T-k_T)^{-N}-(p^\pi_T+k_T)^{-N}}{(p^\pi_T-k_T)^{-N}+(p^\pi_T+k_T)^{-N}} \\
&\simeq N \frac{k_T}{p^\pi_T}
\label{Eq:leftright}
\end{aligned}
\end{equation}
\end{linenomath}
for small  $k_T/p^\pi_T$.
This demonstrates that if the $k_T$ shift is independent of the hard scattering $p_T$,  it is very natural to expect the magnitude of the asymmetry to fall with increasing observed transverse momentum $p_T$ at large $p_T$. 
In previous measurements~\cite{Arsene:2008aa} of the $p_T$ dependence for $A_N$ with charged pions, the asymmetry has been seen to increase with $p_T$ up to about $p_T<1$ GeV/$c$.  In an earlier STAR measurement~\cite{Abelev:2008af}, it was observed that there was little evidence for $A_N$ falling with $p_T$ up to at least 3 GeV/$c$. In this paper, the data are analyzed to separate the independent effects of
$p_T$ and $x_F$.

Two classes of models have been introduced for forward $A_N$, both involved the hard scattering of a leading momentum quark in the polarized proton and both depended upon secondary interactions to generate a spin dependent contribution $\vec{k}_T$ to the pion final state transverse momentum. 
The Sivers effect~\cite{Sivers:1990fh} involved an initial state interaction before the hard scattering of a quark in a polarized proton, leading to initial state parton transverse momentum that depended on the proton transverse polarization. The Collins effect~\cite{Collins:1993kq} generated a transverse spin dependent component to the final state pion transverse momentum from the fragmentation process of the scattered quark, which retained its initial state transverse polarization through the hard scattering process. 
Closely related to Collins and Sivers models was an approach involving higher twist calculation, where the scattered quark was correlated with a soft gluon, which also lead to a significant transverse asymmetry~\cite{Qiu:1998ia}. 

Many model calculations attempt to describe forward pion transverse spin asymmetries using one of these approaches. While for both types of models the basic mechanism involves the production of a final state pion from fragmentation of a hard scattered parton, only the Collins approach explains large $A_N$ arising from the fragmentation process. In contrast to pion production, jet production does not involve fragmentation. The Collins effect therefore does not contribute to that asymmetry. Jet $A_N$ measurements in this kinematic region have been published and the values of $A_N$ were observed to be smaller than measured pion asymmetries~\cite{Bland:2013pkt}. 

Both the Sivers and the Collins approaches introduced a parton transverse momentum relative to the initial or final state hadron momentum to generate a transverse asymmetry without violating helicity conservation.  In the Sivers picture, transverse momentum of initial state quarks can be connected to the initial state orbital angular momentum of a struck quark along the polarization axis. While an orbiting quark does not, on average, have transverse momentum, Sivers argued that absorptive effects could break the left-right symmetric parton $k_T$ distribution to generate the  required non-vanishing average $k_T=\left<\vec{k}_T\cdot \hat{x}\right>$. 
Even though absorption does introduce phase changes, the calculation of this phase in the conventional perturbative calculation was not fully appreciated until it was noted in~\cite{Brodsky:2002rv} that the Wilson line contribution, formally required in the pQCD calculation, did provide exactly the needed phase change for a non-zero $A_N$~\cite{Collins:2002kn}.   

The emerging physical picture is that unlike the case for longitudinal spin dependence,  the observed large values of $A_N$ derive not from the spin dependence of the hard scattering process between the pair of partons, but from the interaction between the scattered quark and the other constituents or fragments of the polarized proton.
While from symmetry, $A_N$ must vanish at $p_T=0$, the example of Eq.~\ref{Eq:leftright} demonstrates that the asymmetry is expected to fall with transverse momentum above some nominal scale, $\vec{k_T}$.
In recent years, there have been many calculations based on Collins, Sivers or twist-3 collinear methods, with a goal to reproduce the basic nature of $A_N$ dependence on kinematics~\cite{ Kouvaris:2006zy, Anselmino:2013rya, Kanazawa:2014dca, Anselmino:2015eoa, Gamberg:2018fwy}.  Within the Collins or Sivers methods it was necessary to account for the longitudinal and transverse momentum distributions of parton momentum within hadrons while traditional collinear parton densities or fragmentation functions involved only longitudinal distributions. In the twist-3 approach, one started with those traditional collinear parton densities or fragmentation functions and dynamically generated the transverse motion from interactions with other fields in the nucleon.   A twist-3 calculation~\cite{Kanazawa:2014dca}, involving fits to many parameters,  resulted in calculations that were in agreement with single inclusive deep inelastic scattering asymmetries and with the $x_F$ dependence of $\pi^0$ $A_N$ in $pp$ collisions. This calculation also resulted in a nearly flat, or very slowly falling, $p_T$ dependence above about $3$ GeV/$c$ for the $\pi^0$ $A_N$ in $pp$ scattering.   While not rising with $p_T$, as do the new $A_N$ $pp$ data presented in this paper, the nearly flat $p_T$ dependence from the twist-3 calculation is interesting. It shows that the intuitive picture of $A_N$ falling with $P_T$, based on the simple arguments of Eq.~\ref{Eq:leftright}, can involve a surprisingly large $k_T$ scale, well above the nominal QCD scale.

\section{Measurements of $A_N$ in Proton-Nucleus Collisions}
If the observed transverse single-spin dependent amplitude for forward pion production arises completely from the localized quark-gluon hard scattering process, then the environment that provides the soft gluon in the second proton or nucleus would not likely impact $A_N$. But we know that the important source of $A_N$ is not the hard quark-gluon scattering process itself but primarily involves the additional interactions with other fields in the nucleon or nucleus, perhaps manifested by the generation of parton transverse momentum relative to the parent hadron momentum.
Because the mechanism responsible for transverse spin asymmetries is not a simple local leading twist interaction but depends on the environment in which a parton interaction occurs, it is clear that $A_N$ could be different for $pp$, $p\rm{Al}$ and $p\rm{Au}$ collisions. Even the simple model of Eq.~\ref{Eq:leftright} reminds us that a change in the shape of the $p_T$ dependence for pion production due to either nuclear absorption, rescattering, or modification of the gluon distribution, could lead to dependence of $A_N$ on nuclear size.

The measurement of how $A_N$ changes when the beam remnant partons of the proton are replaced with spectator partons of a nucleus is a subject of this paper. It is clear that the phase from the Wilson line integral, a line integral of the gauge vector potential color field along the struck quark trajectory, can give rise to color forces between the struck quark and the rest of the polarized proton.   If there are also important color forces between the hard scattering constituents and the residual spectator nucleus, then nuclear dependence of $A_N$ in $pA$ scattering could result. Studies of the spin dependence of the interaction between the interacting quark and the residual spectator nucleus have predicted large nuclear $A$-dependent transverse spin effects but at a lower transverse momentum scale than that of this analysis~\cite{Kovchegov:2012ga}. A more recent calculation was based on lensing forces, with specific reference to the  kinematics of this experiment~\cite{Kovchegov:2020kxg}. The model addressed the dependence of $A_N$ on nuclear saturation as well as the $p_T$ dependence of $A_N$.

One mechanism that could provide nuclear $A$ dependence of $A_N$ relates to the increase in gluon density in the soft gluon distribution probed in forward scattering. It is predicted that at low gluon $x$, when the gluon density becomes large, saturation effects begin to play an important role. For interactions between soft gluons and hard partons producing scattered pions in the range $1.5 <p_T<2.5$ GeV/$c$, saturation effects might modify the interaction, creating significant differences between the corresponding scattering process in $pp$ and $pA$ collisions. Specific saturation models, such as the Color Glass Condensate~\cite{Iancu:2000hn}, predict interactions of the scattered quark with a condensate of gluons rather than a hard scatter from a single gluon.  
Such saturation calculations predict a change in the $p_T$ distribution of the cross section in regions of $p_T$ near the saturation scale, with a suppression of the cross section that increases with nuclear size. In the $p_T\approx 2$ GeV/$c$ range and at more forward pseudo-rapidity than this measurement 
($\eta\approx 4$), STAR has reported that the nuclear modification ratio $R_{d\rm{Au}}$ in $d\rm{Au}$ scattering to produce $\pi^0$ mesons~\cite{Adams:2006uz} is significantly less than unity, suggesting a difference in the scattering process as the size of the nucleus is varied. In the same $p_T$ and rapidity range presented here, measurements of the nuclear modification factors for charged hadrons (mostly charged pions)~\cite{Arsene:2004ux} showed suppression in $R_{d\rm{Au}}$. 
This paper addresses the nuclear dependence of $A_N$, noting, in particular, the lower end of the $p_T$ range where evidence for saturation effects has already been seen in the corresponding $dA$ cross sections~\cite{Adams:2006uz}.

\section{Photon and $\pi^0$ Detection in the FMS}

These data from the Solenoidal Tracker At RHIC (STAR) experiment at the Relativistic Heavy Ion Collider (RHIC) were collected during the 2015 RHIC run, involving collisions between nucleons at center-of-mass energy  $\sqrt{s_{NN}}=200$ GeV per nucleon pair. The photon pair from the decay of the $\pi^0$ was detected with the STAR forward electromagnetic calorimeter, referred to as the Forward Meson Spectrometer (FMS)~\cite{Adam:2018cto}. 
To measure $A_N$ for forward $\pi^0$ production, the STAR detectors used in this analysis were the FMS and the Beam-Beam Counters (BBC).

The two RHIC beams (yellow and blue beams) are 
bunched with up to 120 bunches in each ring. The small angle scattering from the blue beam is associated with positive rapidity.  
Only 111 bunches in each beam are filled and a contiguous set of 9 bunches (the abort gap) are unfilled.
Bunch spacing is 106 ns and the transverse polarization pattern is chosen for each fill according to a predefined pattern (either
alternating the polarization direction from bunch to bunch or for pairs of bunches).
The blue beam  polarization ranged between 50\% and 60\%.

The BBCs are located at a distance of 
$\pm$3.75 meters east and west of the nominal STAR interaction point, 
concentric with the beam line, and covering pseudo-rapidity range
$3.3<\eta<5.2$~\cite{Kiryluk:2005gg,Whitten:2008zz}. 
On both the east and west sides of STAR, each BBC detector consists of an inner and outer hexagonal plane of scintillators.
For heavy-ion collisions, the summed energy deposited in the BBC detectors is related to charged particle multiplicity  
in nucleus-nucleus collisions 
and is sensitive to the event collision centrality. 
As discussed below, for $pA$ collisions we remove events with small signals in the east BBC, on the opposite side to the FMS, to reduce single beam background.

\begin{figure}
\includegraphics[width=0.48 \textwidth]{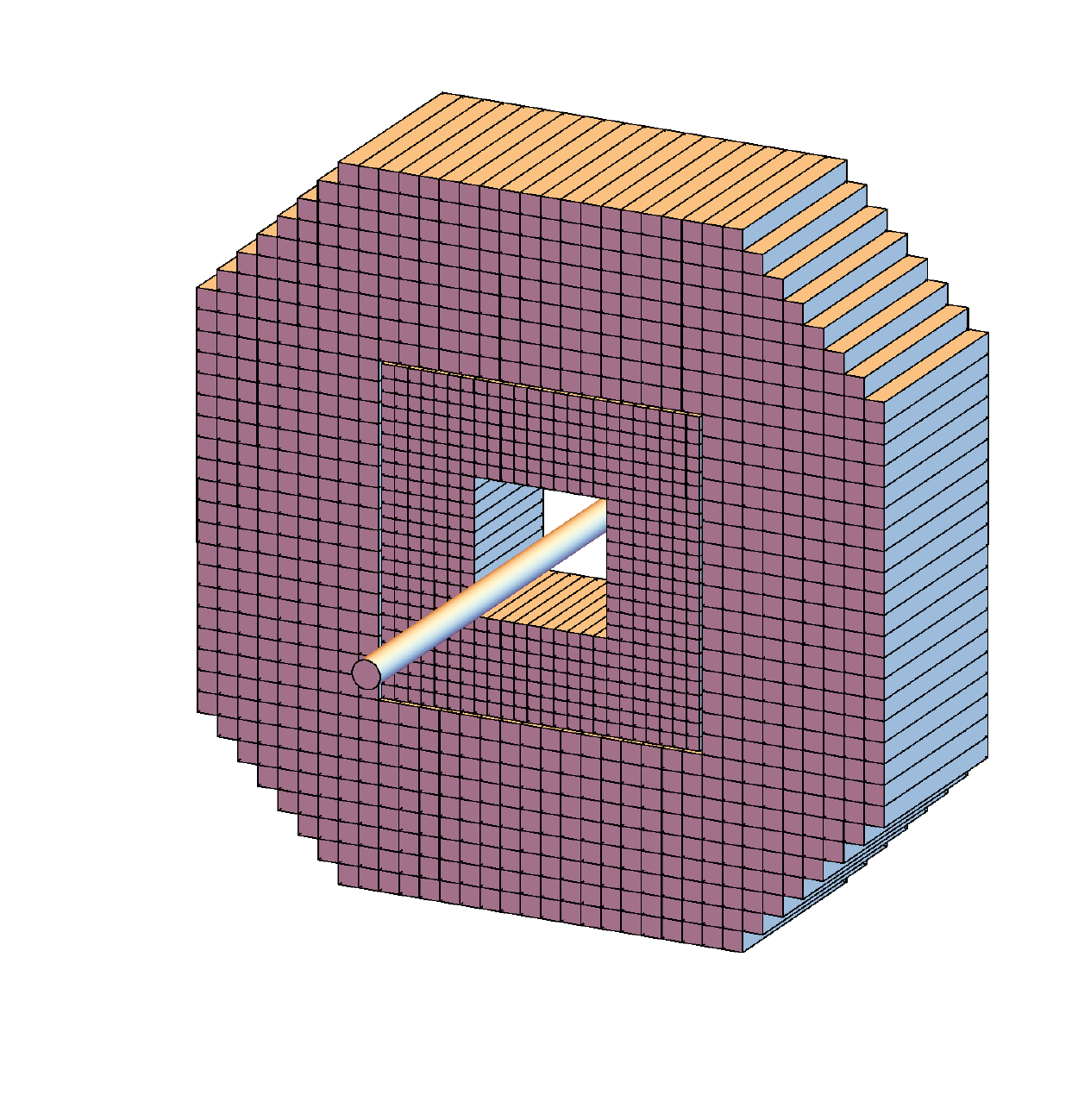}
\caption{\label{fig:FMS} 
The layout for the FMS calorimeter around the RHIC beam-line  located about seven meters west of the nominal STAR interaction point. The FMS consists of lead glass blocks with lengths corresponding to 18 radiation lengths. There are 788 outer blocks with front face dimensions of 5.8 $\times$ 5.8 cm. and 476 inner blocks with front face dimensions of 3.8 $\times$ 3.8 cm.     
}
\end{figure}

The FMS is a Pb-glass electromagnetic calorimeter consisting 
of   1264 rectangular lead glass blocks or cells, stacked in a wall with front 
surface transverse to the STAR beam line as shown in Fig. \ref{fig:FMS}.
The FMS covers the range of forward pseudo-rapidity, $2.7<\eta<3.8$. 
The blocks are of two types, small and large cells.
Details about the detection of $\pi^0$s in the STAR FMS have been discussed elsewhere ~\cite{Adam:2018cto}.

The small and large FMS cells have Pb-glass with different compositions. For small and large cells 
the ratio of cell sizes is chosen to be proportional to the ratio of Moli\'ere radii (transverse electro-magnetic shower dimension); therefore a photon in the large cells will deposit its energy into a similar number of cells as a photon of the same energy in the small cells.
For a 10 GeV photon, the shower distributes measurable energy into about 10 cells. For higher energy photons, the number of involved cells increases. For a 30 GeV photon from the nominal interaction point, incident at the center of a cell, about 80\% of the photon energy is deposited in that cell. Fitting the distribution of energy in cells to an expected distribution from a known shower shape, the transverse coordinates of the incident photon (at shower maximum depth) can be obtained with a resolution of about 20\% of the cell dimension.

In the kinematic range discussed in this paper, observed photons from $\pi^0$ decays have a separation ranging from a few cells to less than one cell. 
For the highest energy $\pi^{0}$s, above 60 GeV, the shower shape from the two photons starts to overlap into a small cell single cluster. Therefore, to reconstruct the highest energy $\pi^0$s, the distribution of deposited energies in cells is fitted to a two photon hypothesis, with parameters that represent the two photon energies and transverse position coordinates. The quality of these fits begins to degrade when the photon separation is on the order of a single cell width.

In addition to photons from $\pi^0$ decays, the FMS measures electrons and positrons. It also has some sensitivity to charged hadrons, such as $\pi^\pm$. On average, a charged pion deposits about 1/3 of its energy in the FMS.  
If the $\pi^0$ is from the fragmentation of a high $p_T$ jet, the FMS sees many of the associated hadronic fragments with degraded energy sensitivity. These charged hadron showers are fit to the photon shower shape and if the deposited energy is greater than 1 GeV, they are included in the list of low energy photon candidates. 
The FMS is triggered by high transverse momentum localized FMS signals. Because these cross sections have a severe transverse momentum dependence, the partially measured charged hadronic background  contributes little to the trigger rate but does contribute background to  $\pi^0$ photon pair signals at high $p_T$.

The events from the FMS where obtained from two trigger methods. The first method is called the board sum trigger, which demands transverse energy to be deposited in localized overlapping rectangles of the 32 FMS cells. The second method is called the jet trigger, which is satisfied by deposition of transverse energy, with a higher threshold than that of the board sum triggers,
measured within overlapping azimuthal regions of angle $\Delta \phi=\pi/2$.  
Three parallel implementations of the board sum triggers are used to select events, each with $\pi^0$ $p_T$ above one of three adjustable thresholds, typically 1.6, 1.9 and 2.2 GeV/$c$. Triggers were prescaled to conserve detector readout bandwidth while sampling the different $p_T$ regions with similar statistical uncertainties. The $pp$ data sample presented in this paper corresponds to an integrated luminosity of 34 pb$^{-1}$ using the highest threshold triggers, which are not prescaled.
The corresponding analyzed luminosity for proton-nucleus 
collisions is 905 nb$^{-1}=\frac{24.5\, {\rm pb}^{-1}}{27}$ and 206 \, nb$^{-1}=\frac{40.6 {\rm pb}^{-1}}{197}$ for $p\rm{Al}$ and $p\rm{Au}$, respectively, where the numerators are provided for direct comparison of proton-nucleon luminosities.

For each event, photon candidates are sorted into ``cone clusters." Each cone cluster includes 
a subset of the photon candidates for which the momentum direction is within an angular cone of 0.08 radians about the cone momentum direction of included photons.
For each photon in the $p_T$ sorted photon event list, the photon is tested for inclusion in the cone cluster list, testing the largest $p_T$ clusters first. If not included in an existing cluster, this becomes the seed of a new cluster.
Usually, only one of these cone clusters will be associated with the large $p_T$ trigger. 
For this analysis of triggered events, only the leading $p_T$ cone cluster is searched for $\pi^0$ candidates. 
This 0.08 radian cone radius, with nominal kinematic pair cuts and for the pion energies around 40 GeV, restricts the selected diphoton mass of photon pairs within a cone cluster to typically less than about 1 GeV/$c^2$. Searching for $\pi^0$ candidates within a cone cluster greatly reduces the combinatorial photon pair possibilities and reduces diphoton background.  

At higher energy, the separation between $\pi^0$ photons becomes small, on the scale of the cell size. In this case, fits to a two photon hypothesis
tend to overestimate the separation between these photons.
For large energy pions, or equivalently large $x_F$, as seen in Fig. \ref{FigMass},
calculated masses are preferentially smeared to larger values.
The $\pi^0$ mass resolution is broadened significantly to higher mass for $\pi^0$ energies $E_{\pi^0}>35~$ GeV in the large cells (lower pseudo-rapidity region) and for energies above $E_{\pi^0}>50$ GeV
in the small cells and higher pseudo-rapidity region
of the FMS.  

The leading energy pair of photons in the highest $p_T$ cluster was analyzed, with selection based on the decay distribution of that two-photon pair. 
The condition $Z < 0.7$ is utilized, where $Z=|\frac{E_1-E_2}{E_1+E_2}|$ and $E_1$ and $E_2$ are the energies of the two photons.
This selection was preferred over a less restrictive one because it decreases background under the $\pi^0$ mass peak. It is the accounting for background under the $\pi^0$ peak that represents the majority of the systematic uncertainty for the measurement of $A_N$.

While it is the intention to measure $A_N$ for inclusive $\pi^0$ production, the selection of the highest-energy two photons for the $\pi^0$ candidates does sacrifice 10-15\% of the pions, depending on kinematics. 
In proton-nucleus collisions ($p\rm{Al}$ and $p\rm{Au}$), we apply an additional selection criterion in order to remove a specific RHIC background which is seen in the “abort-gap” events, between buckets where the nuclear beam is not present. These events are referred to as single-beam events.
For $pA$ collisions, we require that the east BBC have a minimum signal (caused by the breakup of the nuclei). This removes about 5\% of  the lowest activity including most peripheral collisions from this analysis, but also removes nearly all of the single-beam background. 
The residual single-beam background contributes significantly to the systematic error only for a few of the high-$x_F$ bins.

The residual single-beam background fraction in each kinematic bin is estimated from events
seen in the abort-gap bunches.
The ratio of asymmetry for the single-beam background to the $\pi^0$ asymmetry is to be defined as $R_{NB}$, so $A_{NB}=R_{NB} A_N$, where $A_N$ is the $\pi^0$ asymmetry in the particular kinematic bin. Consistent with asymmetries observed in the small number of events in the abort gap, we conservatively assume that $R_{NB}=0.5\pm 0.5$.

\section{The Inclusive $A_N$ Measurements }
\begin{figure*}
\includegraphics[width=\textwidth]{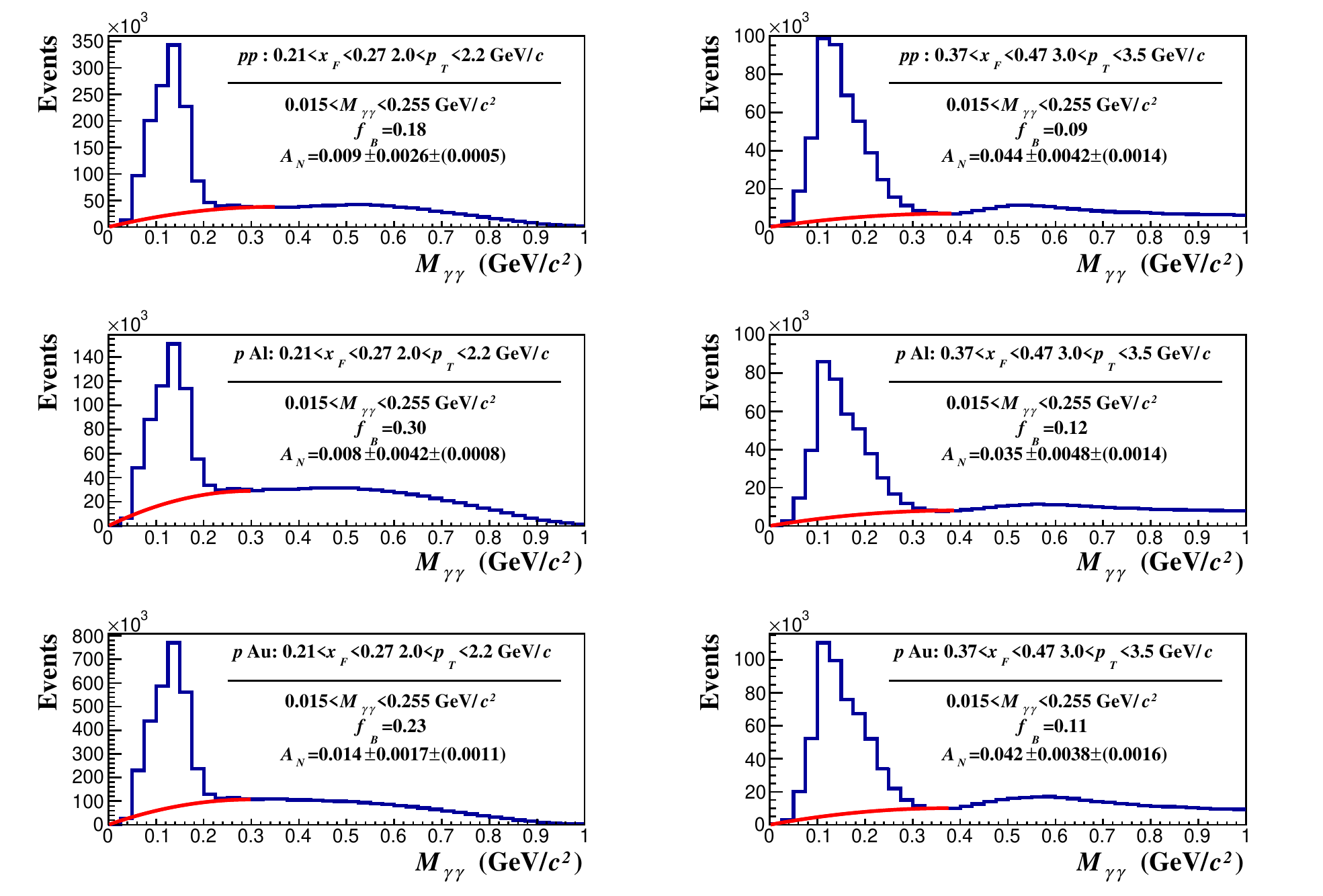}
\caption{\label{FigMass} 
Example invariant mass spectra for diphoton pairs selected within two kinematic regions (two columns) and three collision types (rows: $pp$, $p\rm{Al}$, $p\rm{Au}$).  The asymmetries $A_N$ for pion peaks are obtained within the mass region $0.015<M_{\gamma\gamma}<0.255$ Gev/$c^2$. For the indicated fitted backgrounds under the peaks, the fraction of background events is $f_B$. The measured $A_N$ for the $\pi_0$, with all corrections applied, is included within each panel with statistical uncertainty followed by systematic uncertainty in parentheses.     
}
\end{figure*}
In this paper $A_N$ for forward $\pi^0$ production is measured for $pp$, $p\rm{Al}$, and $p\rm{Au}$ collisions. The high transverse momentum forward $\pi^0$ is detected with the FMS calorimenter, detecting pions with pion pseudo-rapidity $2.7<\eta<3.8$. 
Candidate photon pairs passing the selection are independently analyzed within kinematic regions of $p_T$ and  $x_F$. In Fig. \ref{FigMass}, the diphoton mass, $M_{\gamma\gamma}$, distributions are shown for two example kinematic regions, for $pp$, $p\rm{Al}$ and $p\rm{Au}$ collisions.
The two-photon mass distributions are initially fitted to a quadratic background shape plus a Gaussian pion shape in the mass region below the $\eta$ peak. The Gaussian only approximately represents the shape of the pion peak and that Gaussian shape is only used to determine a mass range above the pion peak.  To finally determine the background fraction, the quadratic background shape is constrained to be zero at a mass of zero and is fit to the mass distribution in the limited mass region above the pion peak. Examples of these background fits are shown in Fig. \ref{FigMass}.
The pion signal is obtained by counting the events in the pion peak, $0.015<M_{\gamma\gamma}<0.255$ GeV/$c^2$, and subtracting the fitted background contribution in that region. 
The typical fraction $f_B$ of background under the pion peak ranges from about 20\% at very low $x_F$ to a few percent when the pion energy is larger. We define $A_B=R_B A_N$ is the asymmetry of the background under the $\pi^0$ peak where $R_{B}$ is the fraction of non-pion background and  $A_N$ is the $\pi^0$ asymmetry.

The value $R_B=0.33 \pm 0.33$ was conservatively determined based on the asymmetry in the mass region ($0.3<M_{\gamma\gamma}<0.4$  GeV/$c^2$) above the pion peak and below the $\eta$ meson peak .  
This background asymmetry cannot be well measured with significance within a single kinematic bin, but is estimated based on an average over many kinematic bins. Uncertainty in this background correction is the most important contribution to the systematic uncertainty in the $\pi^0$ measurement of $A_N$. 
\begin{figure*}
\includegraphics[width=1\textwidth]{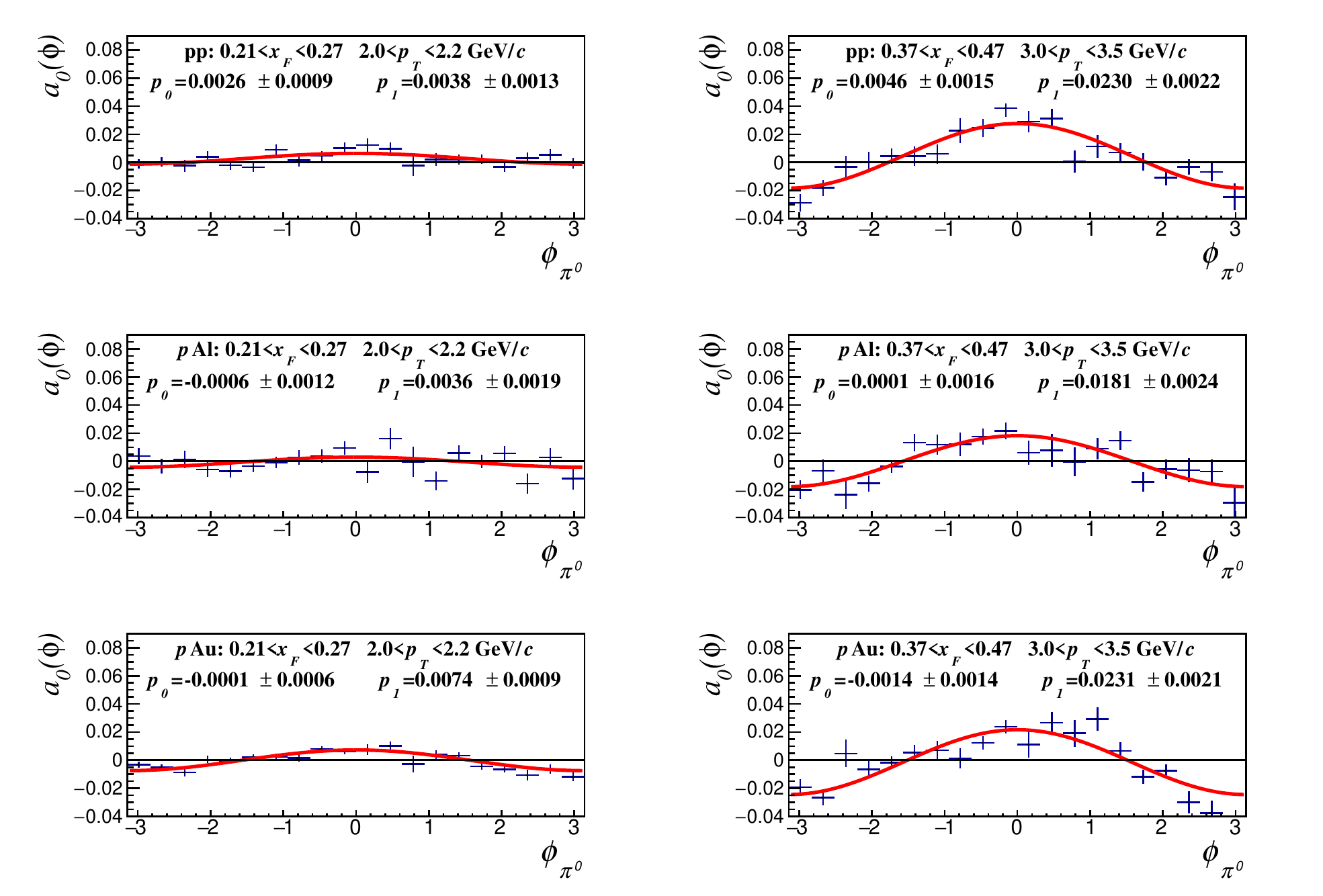}
\caption{\label{FigCos} 
Uncorrected transverse spin asymmetries for the same 6 kinematic regions as in Fig. \ref{FigMass}.  The azimuthal $\phi_{\pi^0}$ distributions of the uncorrected asymmetries, $a_0(\phi)$, are shown for events in the mass range $0.015<M_{\gamma\gamma}<0.255$ GeV/c$^2$. 
Fits to the functional form from  Eq.~\ref{A0Fit} are shown with fitted parameter values $p_0$ and $p_1$. 
}
\end{figure*}
$A_N$ for a given bin in $x_F$ and $p_T$ is extracted from the fits to the uncorrected asymmetries, $a_0(\phi)$, which is determined in each $\phi$ bin from the number of pions ($N^\uparrow$ and $N^\downarrow)$ detected when the proton polarization is up$^\uparrow$/down$^\downarrow$ (see Fig. \ref{FigCos}).
The uncorrected asymmetry is
\begin{linenomath}
\begin{equation}
  a_0(\phi)=\frac{N^{\uparrow}(\phi)-N^{\downarrow}(\phi)}{N^{\uparrow}(\phi)+N^{\downarrow}(\phi)}.
\end{equation}
\end{linenomath}
The azimuthal dependence of $a_0(\phi)$ is fit to the form 
\begin{linenomath}
\begin{equation}
a_0(\phi)=p_0+p_1 \cos\phi.
\label{A0Fit}
\end{equation}
\end{linenomath}
The parameter $p_1$ is proportional to $A_N$ but must be corrected for the polarization of the proton beam $P_B$  and a factor $K$ to account for background effects,
\begin{linenomath}
\begin{equation}
A_N=p_1\frac{K }{P_B}. 
\label{ANdef}
\end{equation}
\end{linenomath}
The beam polarization varied for different RHIC fills. 
The polarization and beam luminosity were largest at the start of a fill and decayed during the fill. To maximize the use of available data acquisition bandwidth, STAR adjusts the FMS trigger prescale factors during the fill, collecting a larger fraction of available low $p_T$ cross section when the luminosity is lower. The analysis of RHIC polarization has been described by the RHIC Polarimetry group~\cite{Schmidke:2017Pol}. 
In this analysis, the average polarization for each kinematic data point is calculated by folding the run by run polarization with the trigger rate contributing to each kinematic point. For a given beam fill, there is variation in the average polarization of  1-2\% for different kinematic regions. 
The variation of $A_N$ from these different polarizations is small with
respect to the overall uncertainties. The uncertainty on polarization
is divided between scale uncertainties common throughout the running
period and non-scale uncertainties that vary fill by fill. The scale
uncertainties, $\Delta P/P$, are 3\%, 3.1\%, and 3.2\% for $pp$, $p\rm{Au}$, and $p\rm{Al}$, respectively,
and are not included in the point-by-point polarization measurement.
When ratios of asymmetries are taken, the dominant polarization uncertainty, like many of the other systematic uncertainties, tends to cancel in the ratio.

In Eq.~\ref{ANdef}, $K$ represents a correction factor to the asymmetry based on the 
estimates of backgrounds in the mass region $0.015<M_{\gamma\gamma}<0.255$ GeV/$c^2$. 
The largest part of the correction $K$ of Eq.~\ref{ANdef} was obtained from the background fraction $f_B$ under the peak with asymmetry $A_{B}=R_B A_N$. The fraction $f_{NB}$ represents a small additional background fraction (typically 1 to 3\%) from interactions that cannot be associated with polarized $pp$ or $pA$ collisions with asymmetry $A_{NB}=R_{NB} A_N$ . Then the factor $K$ is
\begin{linenomath}
\begin{equation}
K=\left[\frac{1}{1+f_B( R_B-1)}\right]\left[\frac{1}{1+f_{NB} (R_{NB}-1)}\right].
\end{equation}
\end{linenomath}

The systematic uncertainties on $A_{N}$ come from three sources: polarization error  (typically $<0.5\%$, excluding the overall polarization scale uncertainty); the beam background (typically $1-30\%$); and the single-beam background (typically $1-3\%$). 
The uncertainty in the multiplicative factor $K$ is the largest source of systematic error in our measurement of $A_N$.
These uncertainties are calculated individually for each given kinematic bin.

The various  systematic contributions to our $p_{T}$ uncertainty have been discussed in detail in a previous analysis~\cite{Adam:2018cto}. The transverse momentum error analysis using that data, collected in 2012 and 2013, is applicable for these 2015 data.  That analysis determined the final $\sigma_{p_{T}}/p_{T}$ to be approximately 5-6\%, an estimate we will adopt here. In both analyses, the dominant contribution lies in the uncertainty on the energy calibration of the detector ($\sigma_{C}\approx5\%$). The energy calibration of the FMS is based on an analysis of the $\pi^0$ mass for 20-30 GeV $\pi^0$ photon pairs in the large cells and 40-50  GeV pairs in the small cells. 
We have conservatively set our final error in transverse momentum, $\sigma_{p_{T}}/p_{T}=7\%$, allowing for minor differences with this analysis and the previous analysis.

The value of the parameter $p_0$ from Eq.~\ref{A0Fit} indicates the asymmetry of relative integrated luminosity, as measured in the given kinematic region.  
RHIC spin patterns are changed for each fill so the integrated luminosities for spin up and spin down bunches are nearly equal.  The distributions of parameters $p_0$ for the three collision system data sets ($pp$, $p\rm{Al}$ and $p\rm{Au}$) have weighted means of ($0.0032\pm0.0002$, $-0.0009\pm0.0002$ and $0.0001\pm0.0002$), respectively. The fits over all kinematic regions to a single constant value, $p_0$, have corresponding $\chi^2$ values of 32, 57 and 45 for 40 kinematic regions (39 degrees of freedom). 
While the extracted values for $A_N$ depend only on the $p_1$ parameter, it is seen from the above that the values of $p_0$ parameters are small and for each beam data set, the measurements of $p_0$ in different kinematic regions are internally consistent within each set.

An $A_N$ point is extracted from each of 110 kinematic and ``collisions beam type" bins based on the value of parameter $p_1$ from the fit to Eq.~\ref{A0Fit}.
As shown for a few example kinematic regions and beam types in Fig. \ref{FigCos}, each two-parameter fit to the 20 azimuthal points results in a $\chi^2$ value.  Over this large ensemble of such fits, the distribution of measured $\chi^2$ values is in good agreement with the theoretical $\chi^2$ distribution. For the $pp$, $p\rm{Al}$ and $p\rm{Au}$ data sets, the average $\chi^2$s for the fits to Eq.~\ref{A0Fit} are 18.5, 18.1 and 18.4 for 18 degrees of freedom, respectively. 
\begin{figure*}
\includegraphics[width=\textwidth]{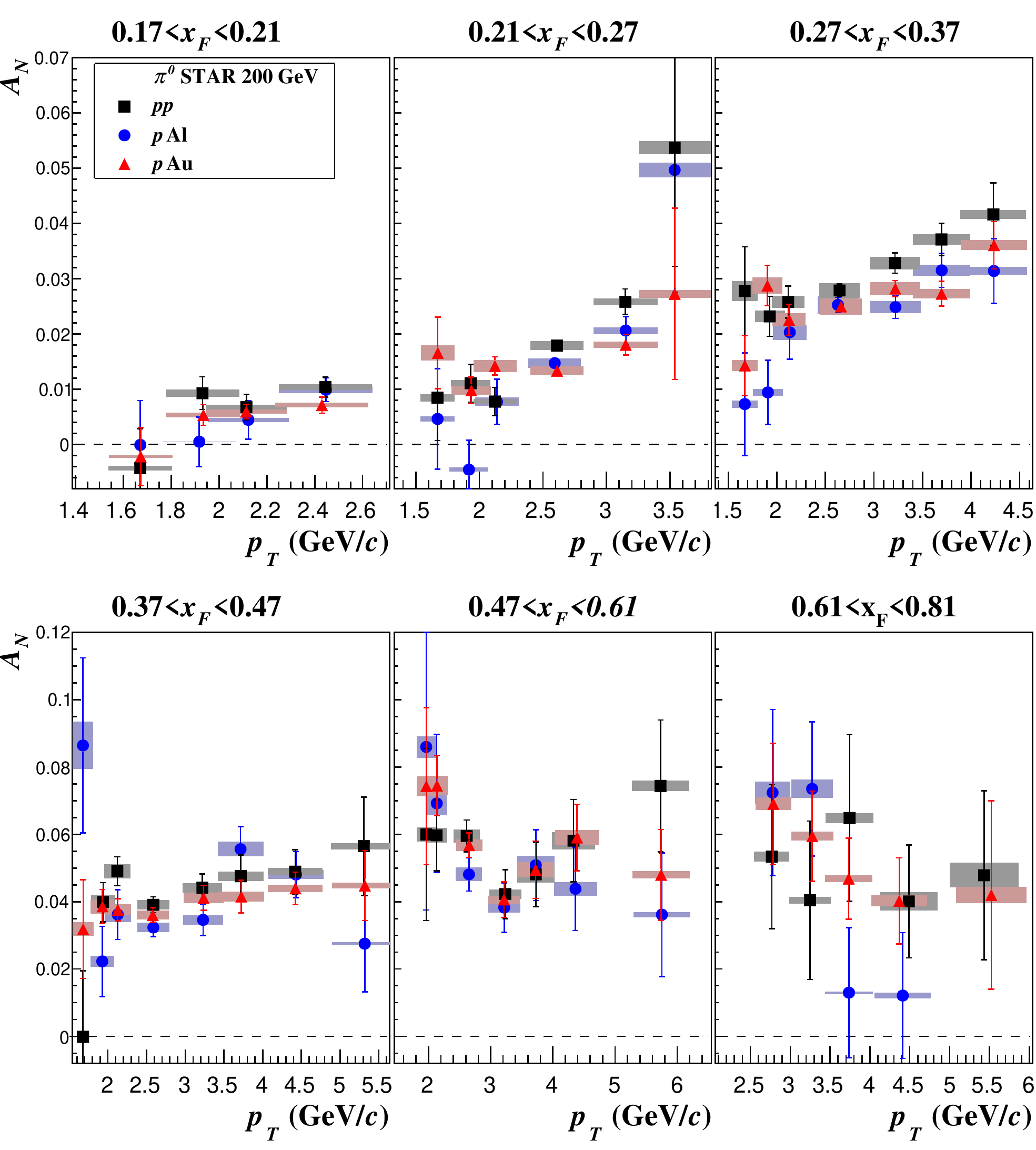}
\caption{ \label{AnPlotsGE2}
The transverse momentum $p_T$ dependence of $A_N$ for 6 bins in Feynman $x_F$. The events contributing are inclusive $\pi^0$s with selection in the invariant mass window $0.015<M_{\gamma\gamma}<0.255$ GeV/c$^2$. Results for the three collision systems are shown, black squares for $pp$, blue circles for $p\rm{Al}$ and red triangles for $p\rm{Au}$ collisions. The event selection criteria are given in the text. The statistical uncertainties are shown with vertical error bars and the filled boxes indicate the horizontal and vertical systematic uncertainties.
}
\end{figure*}
  
\begin{figure*}
\includegraphics[width=\textwidth]{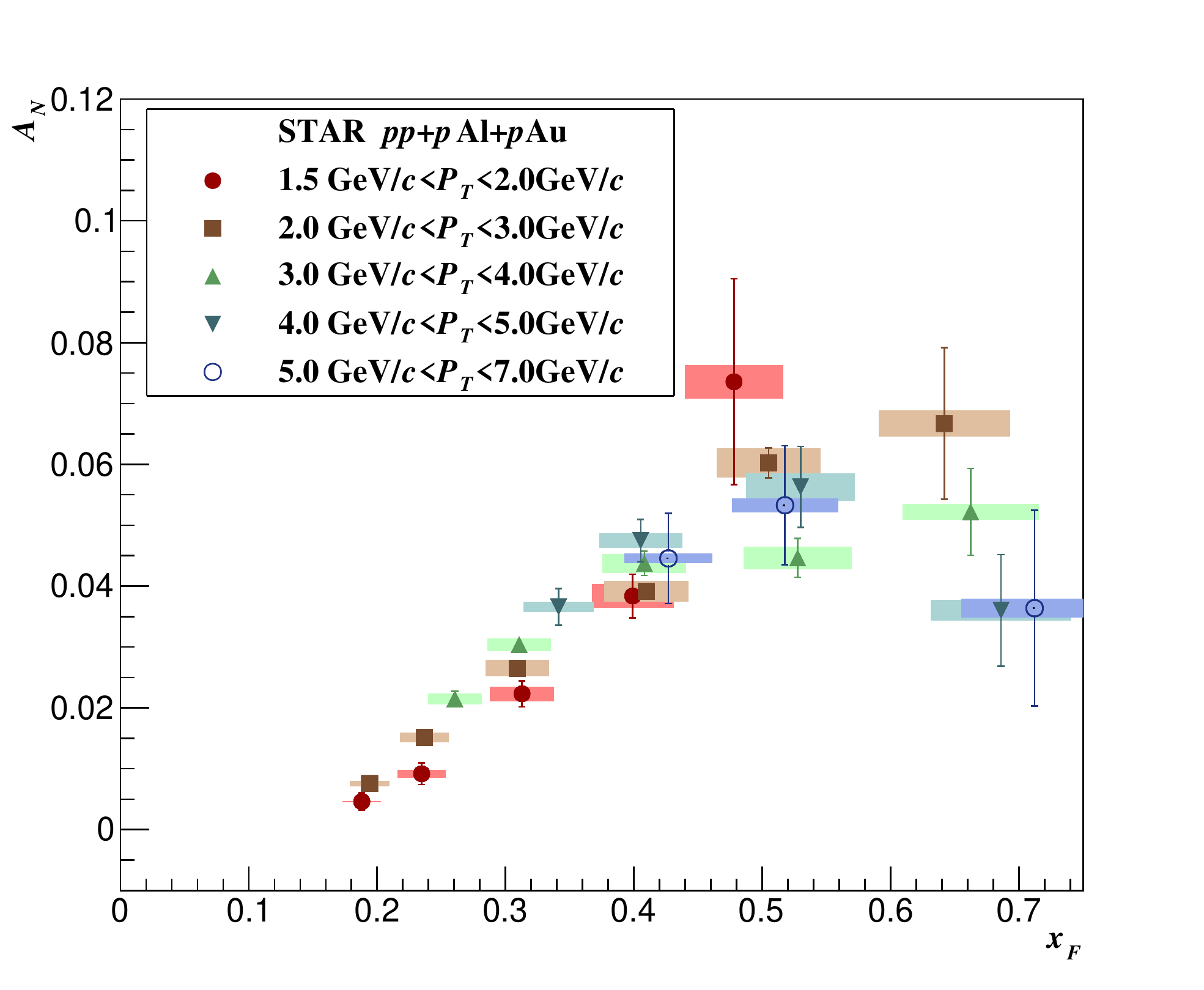}
\caption{\label{FigXFall} 
The $x_F$ dependence of the $\pi^0$ $A_N$ is shown with data from the combined $pp$, $p\rm{Al}$, and $p\rm{Au}$ data points, collecting points within $x_F$ intervals for frames 
from Fig.\ref{AnPlotsGE2} and the indicated $p_T$ range.  Data points  are shown separately for five intervals of transverse momentum indicated by different symbols and plotted horizontally at the average $x_F$ for each combined point. Vertical error bars represent statistical uncertainties and the systematic horizontal and vertical uncertainties are shown with filled boxes.
}
\vspace{.2cm}
\end{figure*}

The examples shown in Figs.~\ref{FigMass} and \ref{FigCos} represent only six kinematic regions of 110 kinematic points at which $A_N$ has been calculated. The transverse single-spin asymmetry for the full data set is shown in Fig. \ref{AnPlotsGE2}. 

Even though $A_N$ is observed to differ among different nuclear collisions systems by 10\% to 20\%, it is an instructive exercise to combine the data sets from different collision systems. 
In Fig. \ref{FigXFall}, the data points from all beams and all transverse momenta are combined in each of the six $x_F$ bins shown in other figures,  
with centers located at $x_F=$ \{0.19, 0.24, 0.32, 0.42, 0.54, 0.71\}. 
All data from  
$pp$, $p\rm{Al}$ and $p\rm{Au}$ collisions are combined and show the $x_F$ dependence for several $p_T$ regions.
For $x_F<0.47$, $A_N$ seems to depend only weakly on transverse momentum, with a gentle increase in asymmetry at larger $p_T$, but at larger $x_F> 0.47$, it appears that $A_N$ may flatten or perhaps falls with $p_T$.

\begin{figure*}
\includegraphics[width=\textwidth]{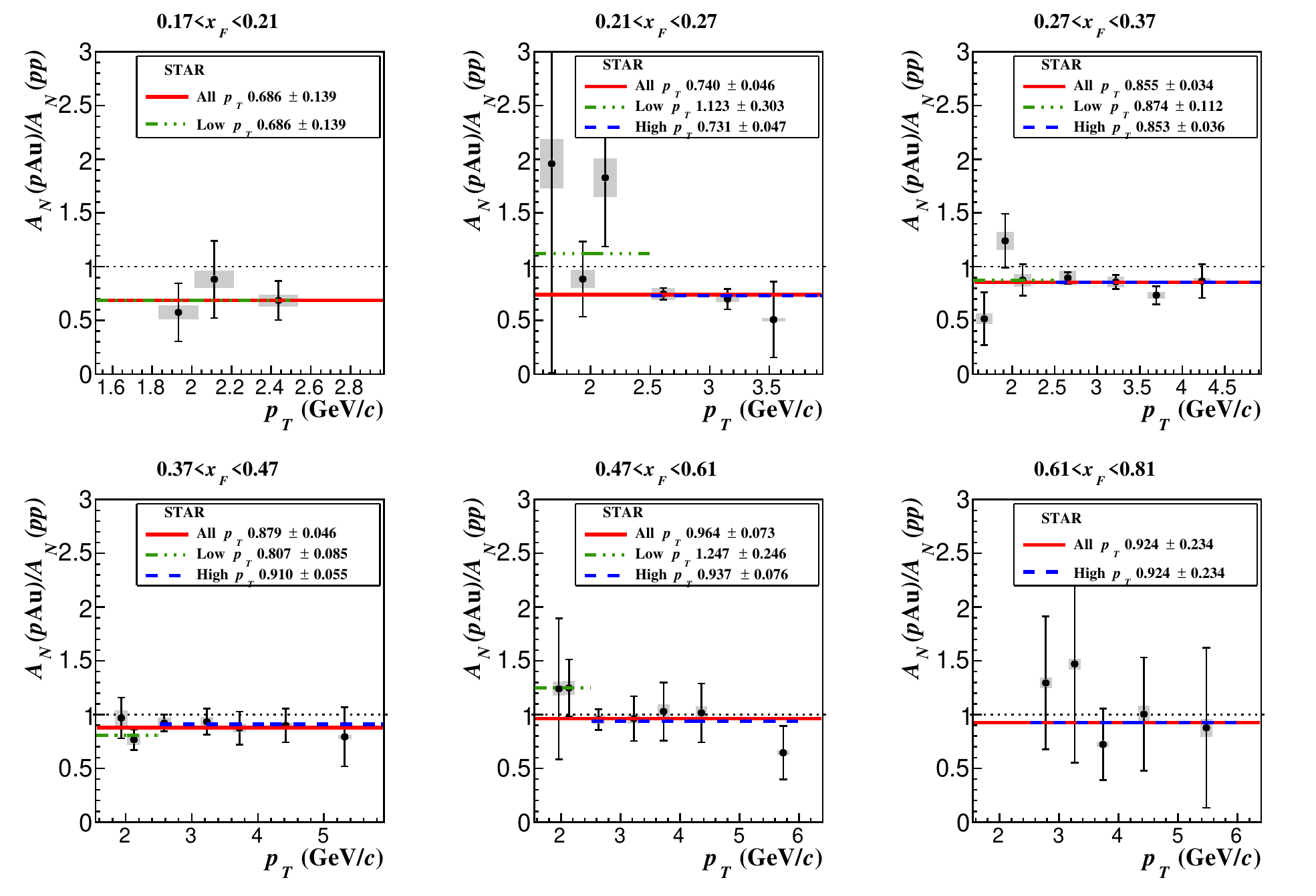}

\caption{ \label{AuppRatAn}
The transverse momentum $p_T$ dependence of the ratio of $A_N$ for $p\rm{Au}$ scattering to that for $pp$ for six Feynman $x_F$ ranges. This figure refers to the same data as is plotted in 
Fig. \ref{AnPlotsGE2}.  The event selection criteria are given in the text. The statistical uncertainties are shown with vertical error bars, and the filled boxes indicate systematic uncertainties appropriate for the ratio.
Horizontal lines indicate the fit to the average ratio over the region $1.5<p_T<2.5$ GeV/$c$, $2.5<p_T<7.5$ GeV/$c$ and for the combined $p_T$ range. 
}
\vspace{.2cm}
\end{figure*}

\begin{figure*}
\includegraphics[width=\textwidth]{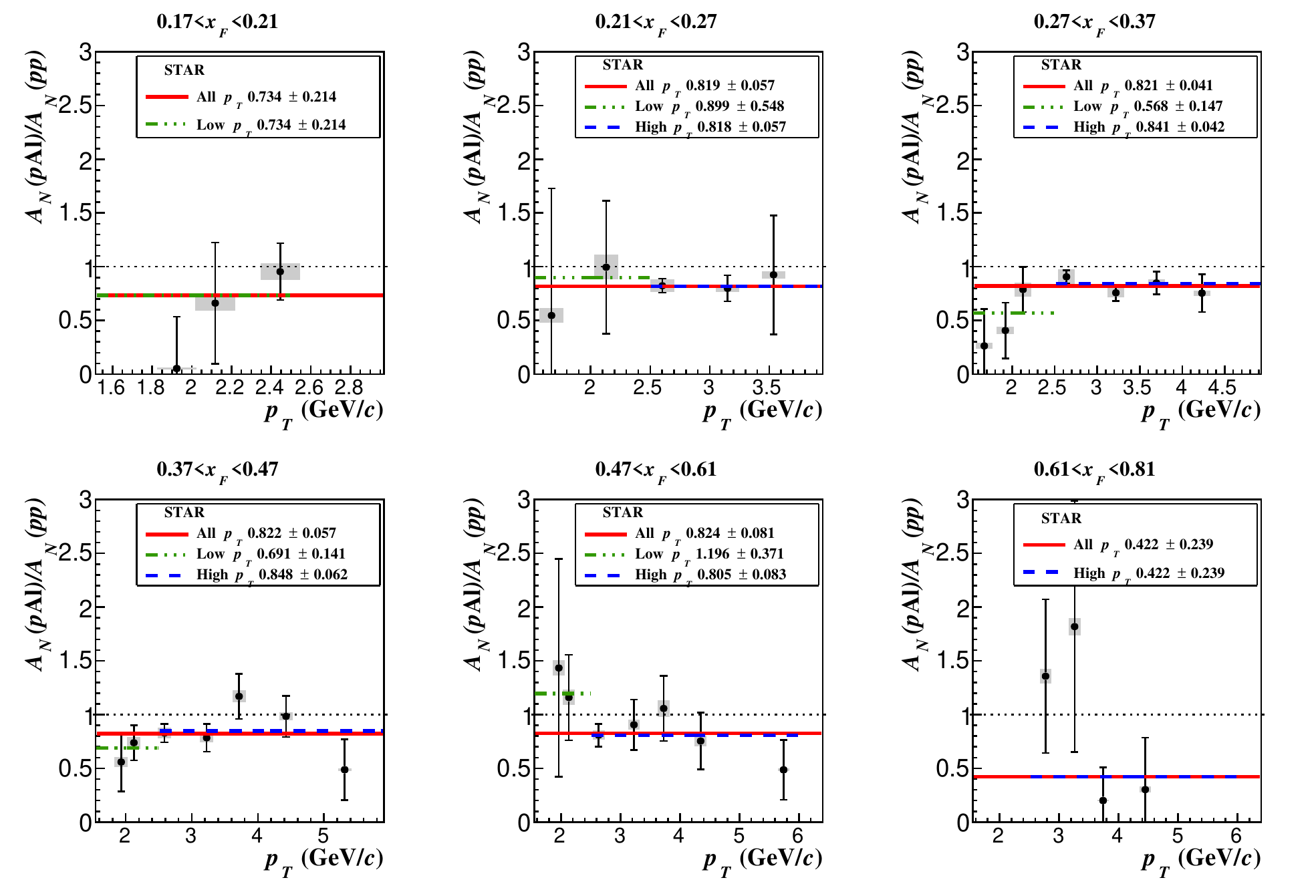}
\caption{\label{RatAnpAlpp}
Similar to Fig. \ref{AuppRatAn} but the ratio of $A_N$ in $p\rm{Al}$ to that in $pp$.  
}
\vspace{.2cm}
\end{figure*}

\begin{figure*}
\includegraphics[width=\textwidth]{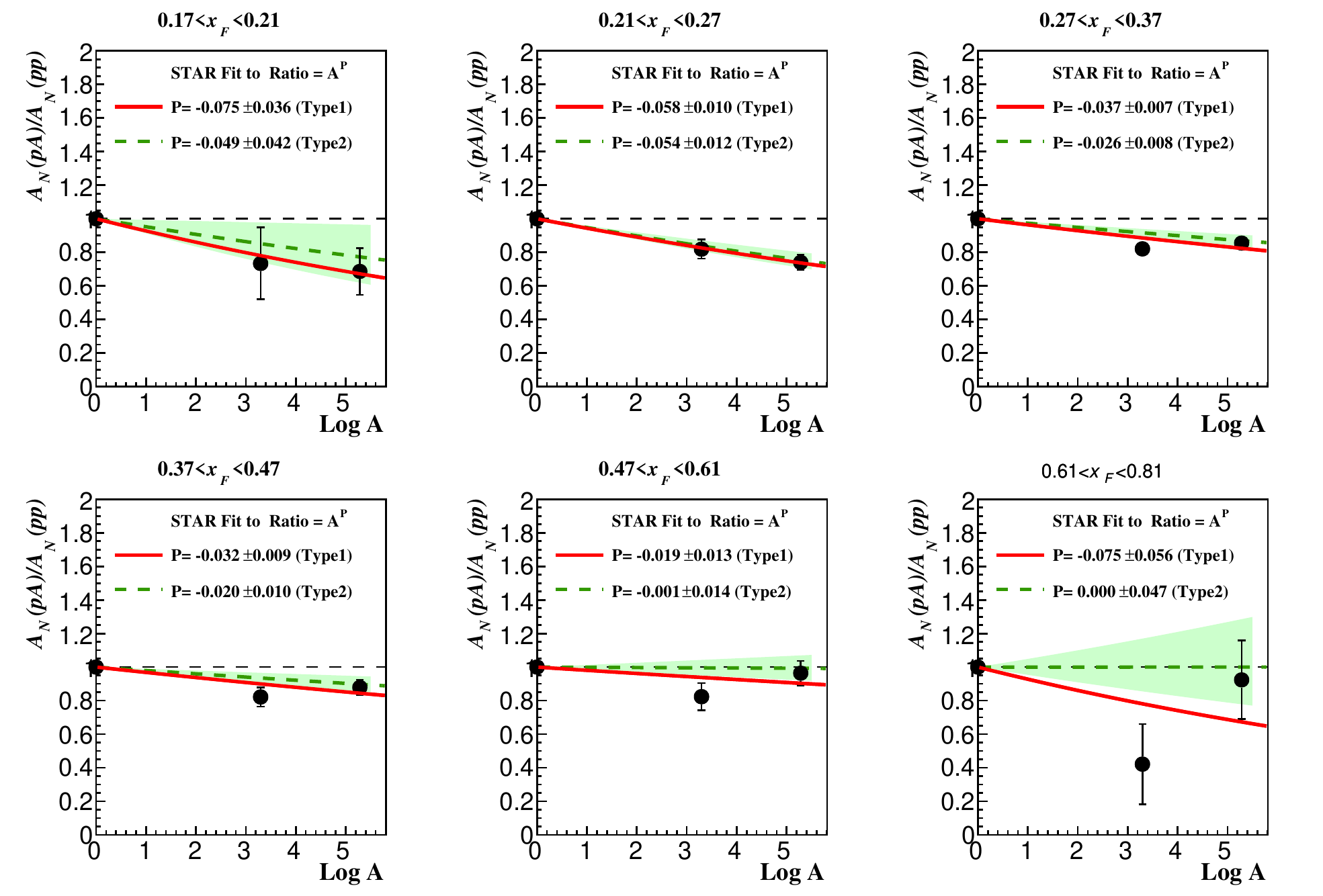}
\caption{\label{rvsAPlots} 
The ratio of $A_N$ for $pA$ scattering to that for $pp$ scattering is shown for six $x_F$ regions, averaging over the full range of $p_T$ dependence. The fitted form for these ratios as a function of $A$ is obtained using Type 1 and Type 2 analyses as described in the text.  The dependence of $A_N$ as a function of $\log A$ is displayed with a filled error band, obtained from the Type 2 analysis, shown as the dashed line.
}
\vspace{.2cm}
\end{figure*}
For each $x_F$ region, the ratios of $A_N$ for $p\rm{Au}$($p\rm{Al}$) to $A_N$ for $pp$ scattering are shown as a function of $p_T$ in Fig. \ref{AuppRatAn}(\ref{RatAnpAlpp}).
The $p_T$ dependences of these ratios are consistent with a constant ratio. Nevertheless, the $A_N$ ratios shown in Fig. \ref{AuppRatAn} and Fig. \ref{RatAnpAlpp} were separately averaged for low $p_T$ ($1.5$ GeV/$c<p_T<2.5$ GeV/$c$) and high $p_T$ 
($p_T>2.5$ GeV/$c$). 
The fitted average values of $A_N$ ratios for each plot in Figs. \ref{AuppRatAn} and \ref{RatAnpAlpp}, averaging over the full $p_T$ range  for each $x_F$, are plotted in Fig. \ref{rvsAPlots} as a function of $\log A$. The systematic uncertainties in  Fig. \ref{AuppRatAn} and Fig. \ref{RatAnpAlpp} are reduced to
account for the correlated background corrections between $pp$ and $pA$ distributions. The non-beam backgrounds thus contribute the most to these systematic errors with statistical uncertainty dominating. 

We parameterize the dependence of $A_N$ on nuclear size $A$ with a power law form
\begin{linenomath}
\begin{equation}
A_N(pA)=A_N(pp)A^P.
\label{eq:AnofA}
\end{equation}
\end{linenomath}

To determine the exponent $P$ for each of the six $x_F$ bins, the weighted means shown in
Fig. \ref{rvsAPlots} are fitted to the power law form,  
\begin{linenomath}
\begin{equation}
\label{eq:rvaAfit}
r(pA)=\left<\frac{A_N(pA)}{A_N(pp)}\right>_{all~p_T}=A^{P}.
\end{equation}
\end{linenomath}
The ratios, $r(pA)$, as defined in Eq.~\ref{eq:rvaAfit}, represent the ratio of nuclear suppression of $A_N$ in $pA$ to $A_N$ observed $pp$ scattering, averaged over the full observed  $p_T$ range.
For each region of $x_F$, we fit to a power law in nuclear size $A$ with a fitted exponent, $P$. Recognizing that the uncertainties in the ratio of $pA$ to $pp$ are correlated, the simple
$\chi^2$ fit in the figure can be biased in the determination of the exponent, $P$. We refer to this simple fit, with correlated uncertainties in the ratios, as a ``Type 1" determination of $P$.

A second method for determining the exponent, $P$, without correlated uncertainties is to fit each point in $p_T$ and $x_F$ to the two-parameter form of Eq.~\ref{eq:AnofA}, 
with parameters $A_N(pp)$ and $P$. 
These fits are two-parameter fits to three measurements within each kinematic region. Then with a weighted mean over $p_T$ of the exponents from fits, an average $P$ is obtained for each $x_F$ region. 
This is referred to as the ``Type 2" method, and the bands corresponding to the one sigma uncertainties in this ``Type 2" fit are shown in Fig. \ref{rvsAPlots} as the shaded regions.

Fitting the exponent of the $A$ dependence of the ratios separately for the low and high $p_T$ regions, the exponents $P_L$ and $P_H$ are obtained,
\begin{linenomath}
 \begin{align}
\label{eq:rvaAfit2}
r_L(pA)=\left<\frac{A_N(pA)}{A_N(pp)}\right>_{p_T<2.5~\rm{Gev}/c}=A^{P_L}\\
r_H(pA)=\left<\frac{A_N(pA)}{A_N(pp)}\right>_{p_T>2.5~\rm{GeV}/c}=A^{P_H}.
\end{align}
\end{linenomath}

Calculations of $A_N$ ratios
by Hatta et al.~\cite{Hatta:2016khv} identify an amplitude
that is thought to be dominant in the saturation region and would scale as $A_N \propto A^{-\frac{1}{3}}$ in $p^\uparrow+A\rightarrow \pi^0 X$. 
These calculations could apply to our present measurements of $A_N$ for $pp$, $p\rm{Al}$ and $p\rm{Au}$ in the transverse momentum range $1.5<p_T<2.5$ GeV/$c$.

\begin{figure*}
\includegraphics[width=\textwidth]{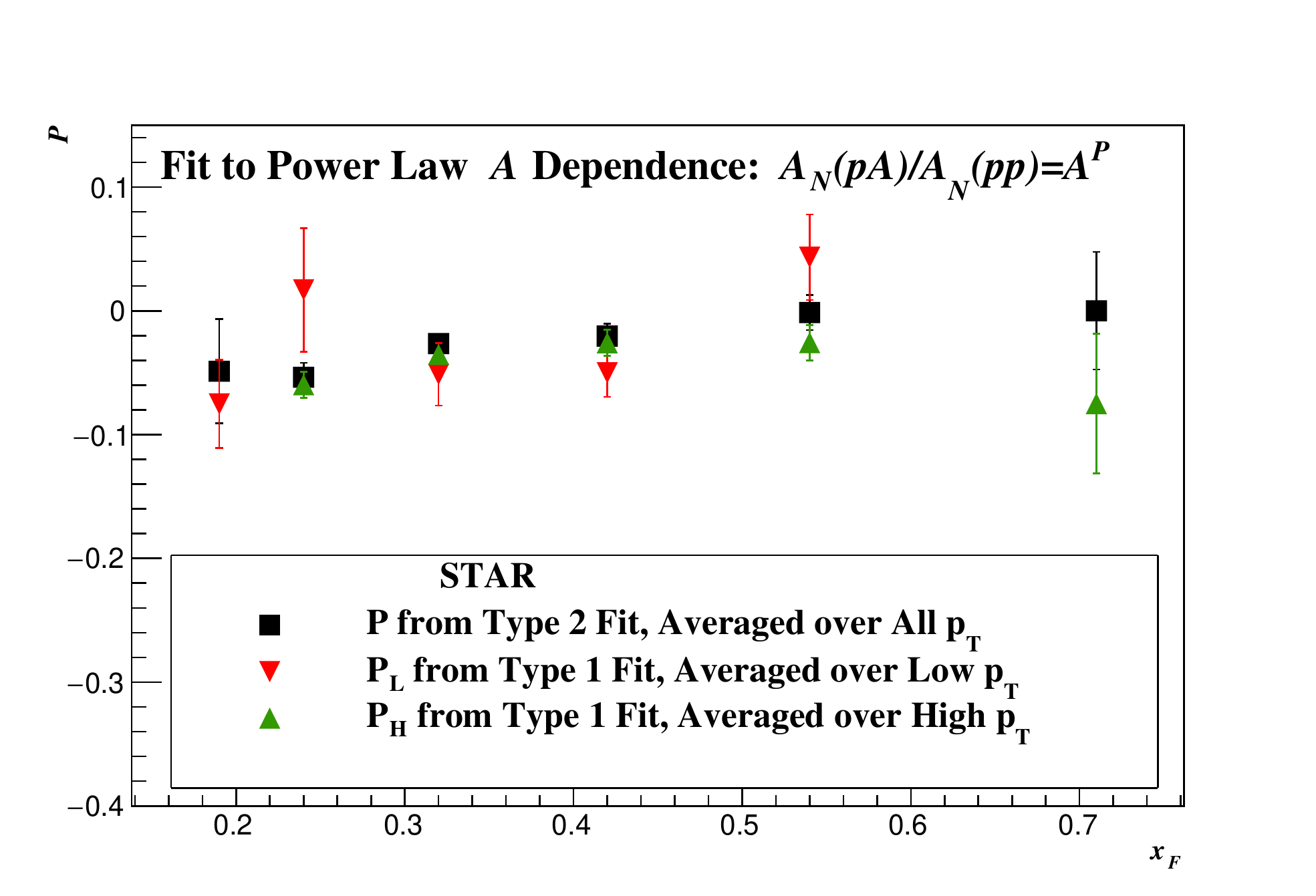}
\caption{\label{rvsAPlots2} 
Analyzing separately the low $p_T$ ($1.5<p_T<2.5$ GeV/$c$) data and the higher $p_T$ data, the exponent, $P$, for nuclear $A$ dependence of the asymmetry ratio $\frac{A_N(pA)}{A_N(pp)}=A^P$ is shown as a function of  $x_F$. Points are included, averaging over the low $p_T$ region and high $p_T$ regions ($p_T>2.5$ GeV/$c$) separately. Examples of one parameter power law fits for $P$ are shown in Fig. \ref{rvsAPlots}, where power dependence exponent $P$ is plotted as a function of Feynman $x_F$. The uncertainties shown are from fits described in the text and are dominated by statistical uncertainties. The systematic uncertainties are small, mostly cancelling,  in ratios between different nuclear $A$ data sets and are not separately shown.  
}
\vspace{.2cm}
\end{figure*}

Comparing gold with A=197 and proton collisions with A=1, this implies a reduction of $A_N$ for $p\rm{Au}$ by more than a factor of 5. Above the saturation region, they predict the $A_N$ will scale as $A^0$, indicating that the transverse single-spin asymmetry at larger $p_T$ could be similar for $pp$ and $p\rm{Au}$ collisions.
The fitted values of the exponents $P_L$ and $P_H$ as functions of $x_F$ are shown 
in Fig. \ref{rvsAPlots2}. The exponents are generally within about 5\% of zero in both the low and high $p_T$ regions and significantly different from the value of $-\frac{1}{3}$ that has been predicted to apply in the region below the saturation scale. 

Another approach~\cite{Kang:2011ni}, based on a geometrical scaling of gluon distributions and with Collins-type fragmentation, has also been used to calculate the transverse single-spin asymmetry. They predicted that for pion transverse momentum below the saturation scale, $p_T^2<<Q_s^2$, the $A_N$ ratio is $A_N(pA)/A_N(pp) \simeq\frac{Q^2_{sp}}{Q^2_{sA}}$, where $Q^2_{sA}$ is the square of the saturation scale for a nucleus with $A$ nucleons. For $p_T$ well above the  saturation scale, the ratio was expected to be 1.  
Models, which suggest that at large $p_T$ the ratio should approach a form with exponent zero, are in good agreement with these data.

\section{Isolated $A_N$ Measurements }

It is observed here that the presence of soft photons or hadronic fragments in the vicinity of the highest $p_T$ pion can decrease the asymmetry significantly, cutting $A_N$ in half in most kinematic regions.
For a subset of the events shown in Fig. \ref{AnPlotsGE2}, there are exactly two photons with energy greater than 1 GeV in the 0.08 radian cone around the $\pi^0$ event. 
We refer to ``isolated" events as those with a highest $p_T$ cone cluster with only a single pair of photon candidates. 
``Non-isolated" events are more jet-like, having at least three photon candidates within the cone.
For a large fraction of the covered kinematics,
about 1/3 of the inclusively selected $\pi^0$ events contributing to Fig. \ref{AnPlotsGE2} have an isolated $\pi^0$. These more exclusive events have generally larger values of $A_N$. 
 
\begin{figure*}
\includegraphics[width=\textwidth]{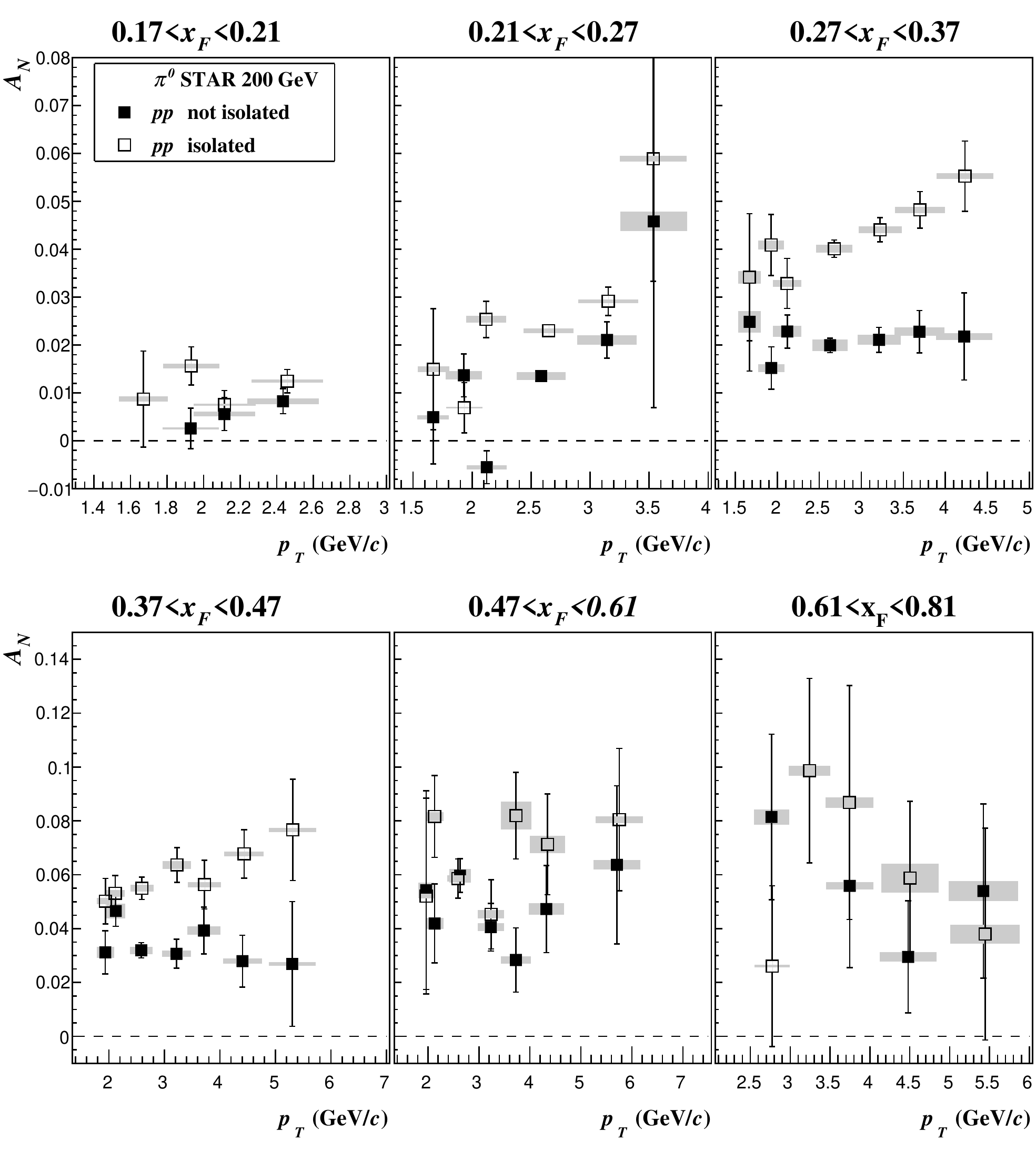}
\caption{\label{AnPlotsE2p} 
The transverse momentum $p_T$ dependence of $A_N$ for pion production in six ($x_F$) regions for $pp$ collisions. The data from Fig. \ref{AnPlotsGE2} have been divided into two parts based on whether 
the $\pi^0$ is produced 
with additional  jet-like fragments of energy more than 1 GeV, shown with filled markers,
or in isolation shown with open markers.  
The event selection criteria for isolated and non-isolated events are given in the text. The statistical uncertainties are shown with vertical error bars. The filled boxes indicate horizontal and vertical systematic uncertainties.
}
\end{figure*}

\begin{figure*}
\includegraphics[width=\textwidth]{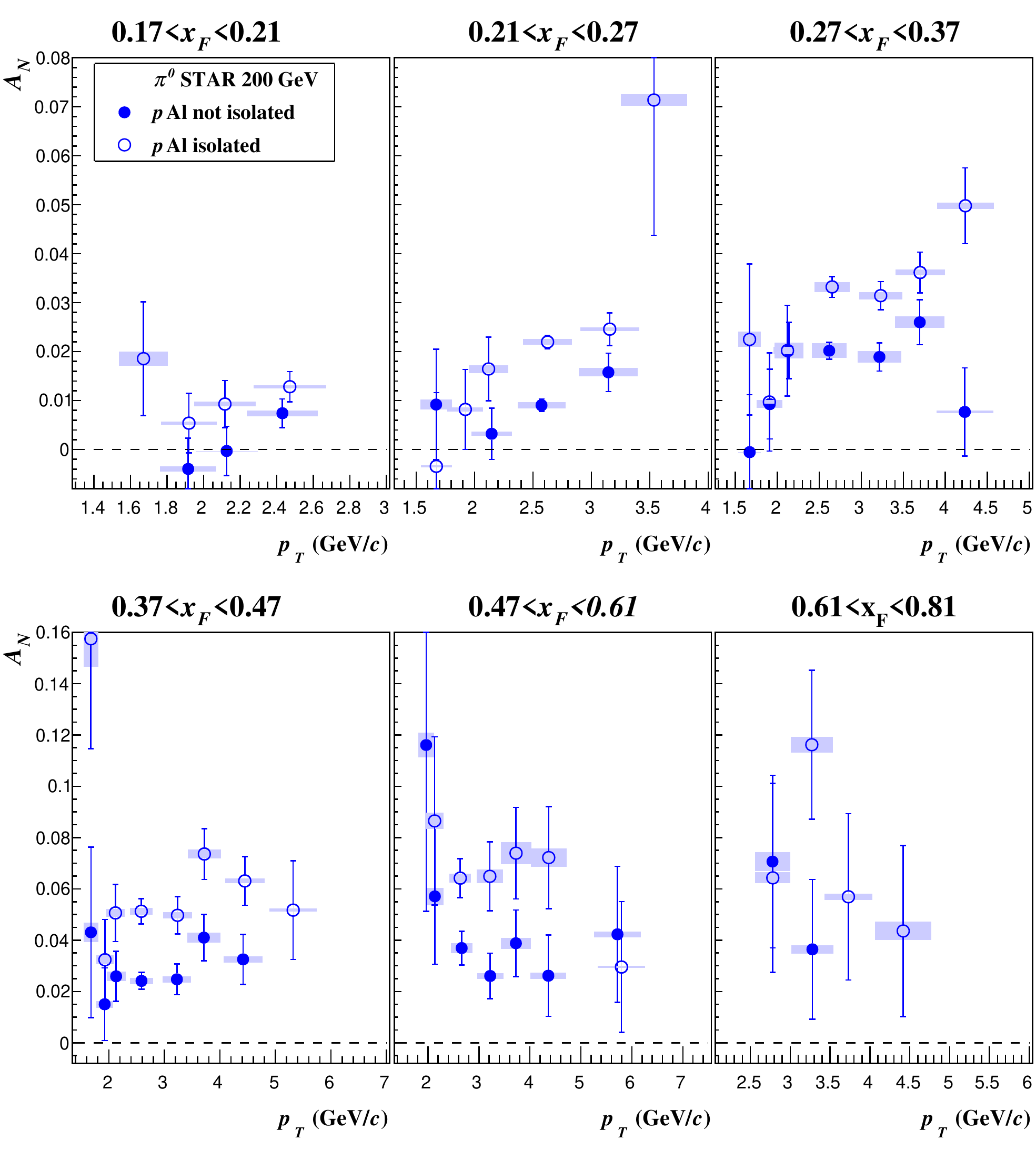}
\caption{\label{AnPlotsE2Al} 
This plot is similar to Fig. \ref{AnPlotsE2p} but for $p\rm{Al}$ collisions.
}
\end{figure*}

\begin{figure*}
\includegraphics[width=\textwidth]{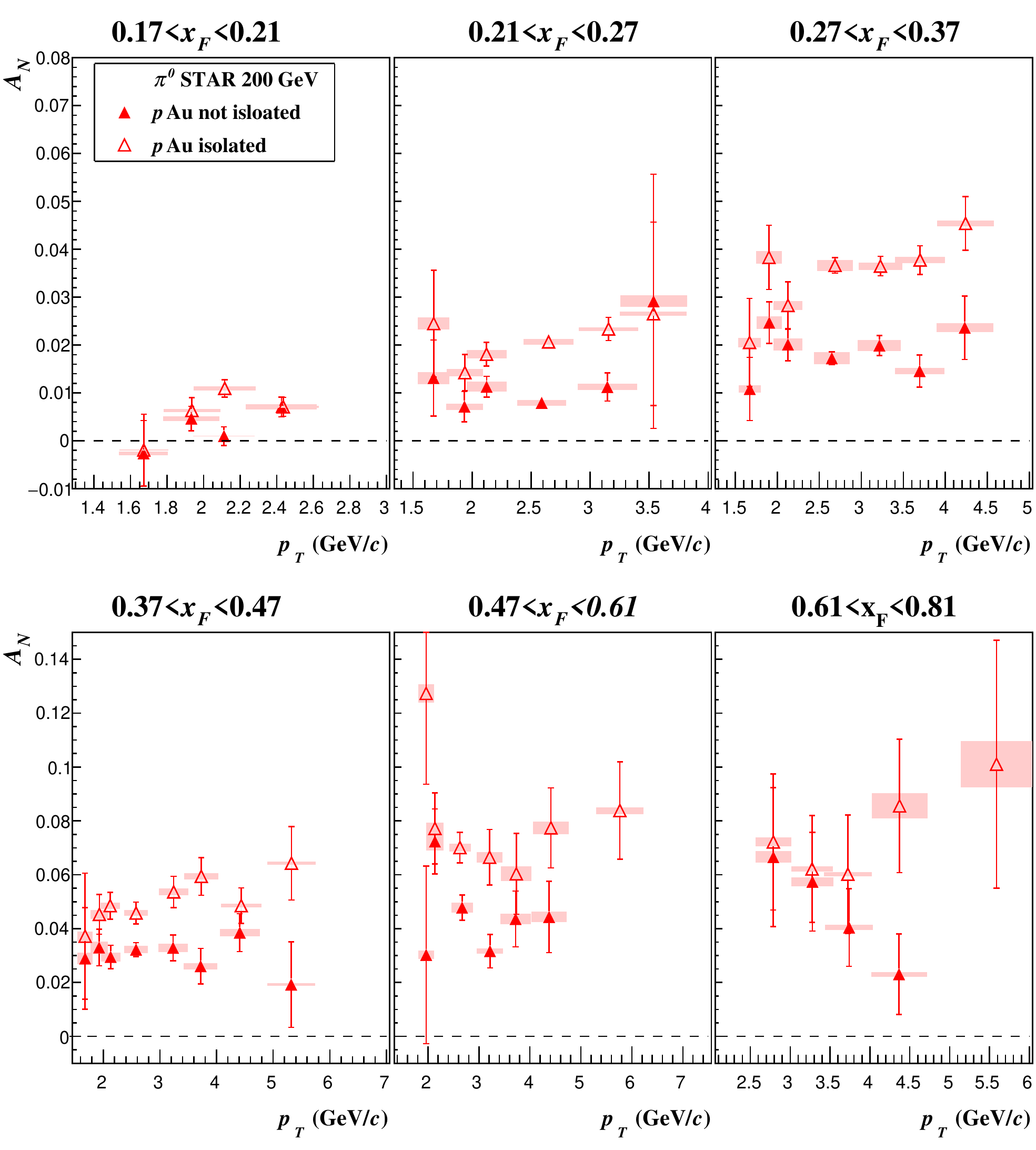}
\caption{\label{AnPlotsE2Au} 
This plot is similar to Fig. \ref{AnPlotsE2p} but for $p\rm{Au}$ collisions.
}
\end{figure*}

It is seen from the comparison of Fig. \ref{AnPlotsGE2} with Figs. \ref{AnPlotsE2p},  \ref{AnPlotsE2Al}, and \ref{AnPlotsE2Au} that $A_N$ for isolated $\pi^0$s is significantly greater than for the complementary part of the inclusive event set with additional fragments observed. 

The electromagnetic calorimeter has limited sensitivity to charged pions, so isolation does not guarantee the absence of hadrons other than $\pi^0$s.  However, this observation hints at the possibility that  the asymmetry for jets with a leading energy $\pi^0$ is much less than the single $\pi^0$ asymmetry in this forward kinematic region. The enhanced $A_N$ for events with no observed jet fragment may indicate that these events are not related to jet production with fragmentation.  

\begin{figure*}
\includegraphics[width=\textwidth]{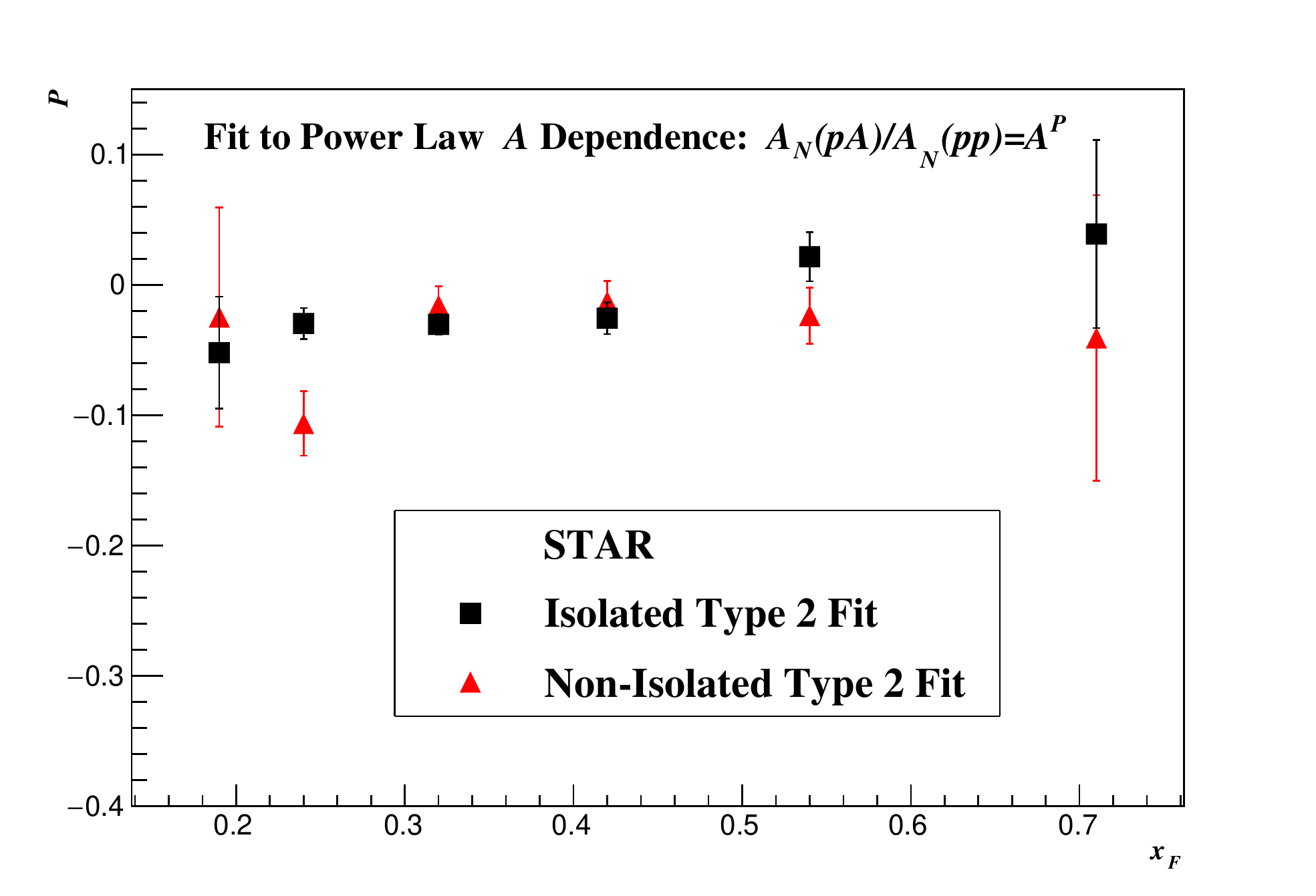}
\caption{\label{PvsXIsoNIso} 
Comparison of the nuclear $A$ dependence of $A_N$ for events with isolated $\pi^0$s and events with non-isolated $\pi^0$s.  
The exponent $P$ of the nuclear $A$ dependence is shown as a function of $x_F$. The exponent is defined in Eq.~\ref{eq:AnofA}. The points shown are averaged over the full $p_T$ range with a Type 2 fit at each $p_T$ and $x_F$ similar to the points of Fig. \ref{rvsAPlots2}.
}
\end{figure*}

The observation that isolated $\pi^0$ events have larger $A_N$ does not appear to depend upon the nuclear size $A$ in $pA$ collisions. In Fig. \ref{PvsXIsoNIso} the determination of the exponent $P$ in the $A$ dependence, defined in Eq.~\ref{eq:AnofA}, has been analyzed separately for isolated and non-isolated events. The average exponents are similar for these two subsets of the data.

This dependence of the measured $A_N$ on event topology is further described in a jet analysis~\cite{Adams:temp}, with some of these same data. Although technical aspects of that analyses differ from this one, the results are consistent in those cases where the same quantity is measured.


\section{Conclusions}
This new measurement of $A_N$ for forward $\pi^0$ production, 
in $pp$, $p\rm{Al}$ and $p\rm{Au}$ collisions, 
determines the dependence on $x_F$ and $p_T$.  
It is observed that $A_N$ generally increases with increeasing $p_T$ at fixed $x_F$ ($0.17<x_F<0.47$), for $p_T$ up to $5~\rm{GeV}/c$.  
In many calculations, 
exemplified by the simple model of Eq.~\ref{Eq:leftright}, 
$A_N$ is expected to fall with $p_T$ when $p_T$ is significantly 
larger than some nominal QCD scale $k_T$, 
representing the spin dependent part of the transverse momentum shift due to
initial or final interactions.
The persistent rise in $A_N$ for $p_T$ well beyond the  $1~\rm{GeV}/c$ scale, is unexpected.

Furthermore, the asymmetry $A_N$, for forward $\pi^0$ production is significantly larger for events with an observed isolated $\pi^0$ than for events that show evidence of additional fragmentation products. It is interesting to compare this result to the published $A_N$ for jets, from~\cite{Bland:2013pkt}, where the asymmetry was observed to be small compared to this $\pi^0$ measurement. 
The Sivers picture, where a proton spin dependent transverse momentum $k_T$ is acquired from initial state interactions, is not the natural choice for explaining the difference in $A_N$ for isolated and non-isolated $\pi^0$s in the final state.
But neither is the enhancement of $A_N$ for isolated pions expected in the Collins picture, where jet fragmentation into multiple hadrons imparts a spin dependent momentum $k_T$ to the observed pion, to generate pion asymmetry. 

The kinematic dependence of $A_N$ on $x_F$ and $p_T$ is similar for the three collision systems.
The suppression of $A_N$ in collisions with nuclear beams is modest, with the typical $A_N$ ratios between $p\rm{Au}$ and $pp$  greater than 80\%. 
When the suppression of $A_N$ is fit to a power law nuclear $A$ dependence, 
$A_N(A)\propto A^P$, the measured exponents from Type 2 fits are in the range of $-0.075<P<0.00$. The weighted average exponent in Fig. \ref{rvsAPlots2} is 
${\langle}P{\rangle}=-0.027 \pm 0.005$. 
This corresponds to a reduction of $r(\rm{Au})= 0.87 \pm0.02$.  
For the Type 1 fits in the low $p_T$ region,
the  weighted average is ${\langle}P_L{\rangle}=-0.037 \pm 0.013$, implying $r_L(\rm{Au})\simeq 0.82 \pm 0.06$.
In the high $p_T$ region, 
the  weighted average is ${\langle}P_H{\rangle} = -0.039 \pm 0.0048$, implying $r_H(\rm{Au})\simeq 0.81 \pm 0.02$.  
There is no significant difference between the exponent $P_H$ in the higher $p_T$ region and $P_L$ in the low $p_T$ region, where gluon saturation effects could be most relevant.  
The general agreement between Type 1 and Type 2 fits helps to give confidence in the fitting methods. 

This nuclear suppression of $\pi^0$ $A_N$ is much less than that reported by the PHENIX collaboration, for positively charged hadrons at somewhat lower pseudo-rapidity or lower $x_F$. The fits from the PHENIX measurement favored an exponent $P=-0.37$~\cite{Aidala:2019ctp}. Unlike the result of this paper, the PHENIX results are nominally consistent with the prediction of Hatta et al., ($P=-1/3$). 

It is noted that the range of $x_F$ coverage by the PHENIX measurement, $0.1<x_F<0.2$, is below the range presented here, shown in Fig. \ref{rvsAPlots2}.
The range of gluon momentum fractions, $x$, probed within the unpolarized beams in this measurement is $x<0.005$, below the $x$ range probed in the PHENIX measurement.   
The distribution of exponents shown in Fig. \ref{rvsAPlots2} indicates that the $P$ exponents slowly increase with increasing $x_F$.
The Type 2 data points can be fit to the linear form $P(x_F)=P_0+{P_0'} x_F$,
yielding fitted parameters $P_0=-0.08\pm0.02$ and ${P'}_0=0.14\pm0.05$.  Linear extrapolation of these data into the center of the 
PHENIX acceptance gives an exponent $P(x_F=0.15)=-0.06\pm 0.02$.  
The PHENIX measurement reported a $\chi^2=\pm 1$ confidence range, expressed here as a $P$ range, of approximately $-0.6<P<-0.25$. Comparing this to the STAR extrapolated value, the difference appears significant. From the $\chi^2$ plot in the PHENIX paper, the value, $P=-0.06$, corresponds to $\chi^2\simeq 13$.  Of course, the linear extrapolation is just an assumption.

Combining all beam types to maximize statistics for $A_N$ measurements, 
for Feynman $x_F<0.47$ the asymmetry $A_N$ 
increases with $x_F$ and with $p_T$.  
For $x_F>0.47$ the trend moderates, 
as the dependence of $A_N$ on $p_T$ flattens or may begin to fall with $p_T$ 
over the measured $p_T$ range. 

These measurements of the dependence of $A_N$, for forward $\pi^0$ production, on kinematics and event topology, should provide new input for ongoing theoretical studies of the underlying dynamics for these processes.
In $pA$ collisions, the dependence of $A_N$ on nuclear size $A$ has been measured and is small.

\section*{Acknowledgments}
We thank the RHIC Operations Group and RCF at BNL, the NERSC Center at LBNL, and the Open Science Grid consortium for providing resources and support.  
This work was supported in part by the Office of Nuclear Physics within the U.S. DOE Office of Science, the U.S. National Science Foundation, the Ministry of Education and Science of the Russian Federation, National Natural Science Foundation of China, Chinese Academy of Science, the Ministry of Science and Technology of China and the Chinese Ministry of Education, the Higher Education Sprout Project by Ministry of Education at NCKU, the National Research Foundation of Korea, Czech Science Foundation and Ministry of Education, Youth and Sports of the Czech Republic, Hungarian National Research, Development and Innovation Office, New National Excellency Programme of the Hungarian Ministry of Human Capacities, Department of Atomic Energy and Department of Science and Technology of the Government of India, the National Science Centre of Poland, the Ministry  of Science, Education and Sports of the Republic of Croatia, RosAtom of Russia and German Bundesministerium fur Bildung, Wissenschaft, Forschung and Technologie (BMBF), Helmholtz Association, Ministry of Education, Culture, Sports, Science, and Technology (MEXT) and Japan Society for the Promotion of Science (JSPS).
\clearpage 
\newpage

\renewcommand{\thepage} {\arabic{page}}


\end{document}